\documentclass[useAMS,usenatbib]{mn2e}
\usepackage{natbib}
\usepackage{graphicx}
\usepackage[
bookmarkstype=toc=true,
 linktocpage=true,
 bookmarks=true,
 unicode
 ]{hyperref}
\usepackage{amsmath}
\usepackage{color}
\usepackage[varg]{txfonts}
\usepackage{comment}
\usepackage{xcolor}

\def\bs#1{\mbox{\boldmath $#1$}}

\newcommand{\ajn}[1]{{\color{red}[AJN: #1]}}
\newcommand{\Msun}{{\rm  M_{\odot}}}
\newcommand{\himpc}{{h^{-1}{\rm Mpc}}}
\newcommand{\hmpci}{{h{\rm Mpc}^{-1}}}
\newcommand{\himsun}{{h^{-1}{\Msun}}}
\newcommand{\HI}{\textsc{Hi}}
\newcommand{\val}[3]{$#1_{#2}^{#3}$}
\newcommand{\RA}[1]{{\color{black} #1}}
\newcommand{\rev}[1]{{\color{black} #1}} 

\begin{document}

\title[RSD of \HI\ gas with hydrodynamic simulations]%
{Redshift Space Distortion of 21cm line at $1<z<5$
  with Cosmological Hydrodynamic Simulations}

%
%
\author[Ando et al.]
{Rika Ando$^{1}$,
\thanks{Email: ando.....}
Atsushi  J. Nishizawa$^{1,2}$,
\thanks{Email:atsushi.nishiza@iar.nagoya-u.ac.jp}
Kenji Hasegawa$^{1}$,
Ikkoh Shimizu$^{3}$
and
\newauthor
Kentaro Nagamine$^{3,4,5}$\\
%
%
$^{1}$ Department of Physics, Nagoya University, Furocho, Chikusa, Nagoya, Aichi 464-8602, Japan,\\
$^{2}$ Institute for Advanced Research, Nagoya University, Furosho, Chikusa, Nagoya, Aichi 464-8602, Japan,\\ 
$^{3}$ Department of Earth and Space Science, Graduate School of Science, Osaka University, 1-1 Machikaneyama, Toyonaka, Osaka 560-0043, Japan\\
$^{4}$ Department of Physics and Astronomy, University of Nevada, Las Vegas, 4505 S.Maryland Pkwy, Las Vegas, NV, 89154-4002, USA\\
$^{5}$ Kavli IPMU (WPI), The University of Tokyo, 5-1-5 Kashiwanoha, Kashiwa, Chiba, 277-8583, Japan \\
}
\maketitle
\begin{abstract}
We measure the scale dependence and redshift dependence of 21 cm line emitted from the neutral hydrogen gas at redshift $1<z<5$ using full cosmological hydrodynamic simulations by taking the ratios between the power spectra of \HI--dark matter cross correlation and dark matter auto-correlation. The neutral hydrogen distribution is computed in full cosmological hydrodynamic simulations including star formation and supernova feedback under a uniform ultra-violet background radiation. We find a significant scale dependence of \HI\ bias at $z>3$ on scales of $k\gtrsim 1 \hmpci$, but it is roughly constant at lower redshift $z<3$. The redshift evolution of \HI\ bias is relatively slow compared to that of QSOs at similar redshift range.
We also measure a redshift space distortion (RSD) of \HI\ gas to explore the properties of \HI\ clustering. Fitting to a widely applied theoretical prediction, we find that the constant bias is consistent with that measured directly from the real-space power spectra, and the velocity dispersion is marginally consistent with the linear perturbation prediction. Finally we compare the results obtained from our simulation and the Illustris simulation, and conclude that the detailed astrophysical effects do not affect the scale dependence of \HI\ bias very much, which implies that the cosmological analysis using 21 cm line of \HI\ will be robust against the uncertainties arising from small-scale astrophysical processes such as star formation and supernova feedback. 
\end{abstract}

cosmology: theory -- galaxies: formation  -- radio lines: general -- intergalactic medium --  hydrodynamics

\section{introduction}
\label{sec:introduction}
The acceleration of the Universe has been one of the greatest mysteries since it was first discovered by the observations of type Ia supernovae \citep{Perlmutter+1999}. One of the most natural explanations of the accelerated expansion is the dark energy in the regime of general relativity or modified theory of gravity \citep[e.g.][for review]{Clifton+2012}. Because the acceleration only becomes effective at the late epoch of $z\lesssim 1$, the most promising probe of dark energy or modified gravity is the large-scale structure of the Universe.

Baryon acoustic oscillation (BAO) is recognized as a useful technique which is least affected by the systematics to constrain the dark energy models  \citep[e.g.][]{Albrecht+:2006}. After the first detection of BAO by the clustering of luminous red galaxies (LRG) in the Sloan Digital Sky Survey (SDSS) \citep{Eisenstein+:2005}, 
\rev{\RA{significant attention has been paid to constrain the dark energy using BAO in the power spectrum and correlation function}} \citep[e.g.][]{Ross+2015,Beutler+2011}. 
As the BAO is a measurement of the oscillation peak scales, an accurate prediction of the peak scales is required. It is well known that the oscillation peak scale is readily changed by the non-linear clustering of matter \citep{Nishimichi+2007} or the non-trivial couplings among different fluctuation modes due to galaxy bias \citep[e.g.][]{Cole+:2005, Dalal+2008b, McDonald+:2009}.
Another important aspect of the BAO is the combination of parallel and perpendicular components to the line of sight \citep{AP+:1979}. Although the AP-test makes the BAO a more powerful tool to constrain cosmological parameters, the systematic effect due to redshift space distortion (RSD) has to be taken into account as it is degenerate with the AP effect.

As we can observe galaxies only in redshift space, the distance to the galaxies are contaminated by the peculiar velocities of galaxies; on large scales, galaxies are coherently attracted toward the overdensity regions which makes the anisotropic two dimensional correlation function squashed, while on small scales, non-linear random motion makes correlation function elongated along the line of sight \citep[e.g.][]{Matsubara:2004}. The RSD is important not only for correctly understanding the distortion of the correlation function to utilize the AP effects in the BAO, but also to gain an independent cosmological information from the BAO. Since the RSD is a direct measure of the velocity field, it is sensitive to the potential fluctuation $\Phi$ and thus to the theory of modified gravity \citep[e.g.][for review]{Hamilton:1998}. 

The current measurements of the BAO and RSD have been mainly focused on the galaxy or QSO distribution as they are considered to be good tracers of the large-scale structure \citep{Kirshner+1981, Alam+2017, Padmanabhan+2007, Torre+2017, Zhao+2016, Busca+2013, Slosar+2011}. However, due to the difficulty of taking the spectrum to accurately measure the redshift of the sources, we have studied the BAO only at $z<3$.  The first detection of the reionisation absorption signature by the EDGES observation of 21 cm line absorption associated with neutral hydrogen (\HI) gas \citep{Bowman+:2018} has also opened a new window to probe the large-scale structure. Even after the epoch of reionisation, some fraction of neutral hydrogen is confined within the high-density regions such as inside the galaxies preventing the ultra-violet photons to penetrate. Several surveys to map the 21 cm distribution by intensity mapping are proposed where individual objects are not resolved but a continuous smoothed sky distribution is mapped out.  For example, the Square Kilometre Array (SKA) will cover the 25,000 (SKA1) square degree of sky with 50kHz frequency resolution, which is adequate for accurate redshifts at $0.35<z<3.06$ (for SKA1-MID), but with a moderately coarse angular resolution for the single-dish observation \citep{Santos+:2015, Bull+2015}. There are interferometer mode in SKA which has significantly better angular resolution (depends on the configuration), however, its small field of view is not suited for a wide sky coverage and thus for cosmological analyses. Another example is the Baryon acoustic oscillations In Neutral Gas Hydrogen (BINGO) which will target much lower redshifts at $0.13< z<0.48$ \citep{BINGO:2012}. BINGO will cover $15\times 200$ square degree of sky with 1MHz frequency resolution with a resolution of 40 arcmin for single-dish observation.

Much attention has been paid to the cosmological application of \HI ~observations \citep[e.g.][]{Camera+:2015, Bull+2015, Raccanelli+2015, Olivari+:2018, Villaescusa-Navarro+:2017, Obuljen+:2018, Dinda+:2018}. 
It has been shown that the Square Kilometre Array (SKA) has a capability to constrain the dark energy parameters comparable to that from the galaxy redshift surveys such as Euclid \cite{Bull+2015} which is a Stage-IV survey according to the Dark Energy Task Force \citep{Albrecht+:2006}.   
However, it is also shown that the difficulty of the \HI\ observation lies in the foreground removal \citep[e.g.][]{Wyithe+:2005} and in the modeling of \HI\ bias.
As in the case for galaxies or QSOs, it is important to understand the connection between \HI\ gas and dark matter distribution, as theoretical predictions are often only for dark matter, the most dominant component of matter in terms of gravitational interaction. 
While \cite{Bull+2015} assumed the simplest constant bias, \cite{Umeh:2017} found that the non-linear coupling between different fluctuation modes or shot-noise can modulate the amplitude of power spectrum even on large scales. Therefore, considering more realistic models or measurements from numerical simulations is of great importance for robust cosmological analyses.
To predict the \HI\ bias, halo model approach has been considered \citep{Padmanabhan+:2017, Penin+:2018}, and several work have been done using pure N-body simulations \citep{Sarkar+2016,Sarkar+2018a}. They do not fully solve the radiative transfer but populate the \HI\ according to the mass of the host dark matter halo.

In this paper, we measure the scale- and redshift-dependent \HI\ bias using the full cosmological hydrodynamic simulation developed by the Osaka group, which takes gas dynamics into account with an appropriate UV background radiation and star formation with supernova feedback.  We also perform the same analyses using the publicly released data of the Illustris simulation, and compare with our results to examine the impact of AGN feedback on the cosmological signals. 
\rev{{It is well known that $\Omega_{\rm HI}$ depends on the mass resolution of the simulation \citep[e.g.][]{Nagamine2004a,Dave2013}, therefore we use the Illustris-3 simulation which has  similar resolution to our fiducial run.  This allows us to perform a fair comparison and to minimize the effect of mass resolution. }}

This paper is organized as follows. In Section~\ref{sec:simulation}, we describe the details of cosmological hydrodynamic simulations used in this work, 
and explain how we generate the mock \HI\ data based on these simulations.
In Section~\ref{sec:pspec}, we show the results for measuring the \HI\ bias in real space to explore the redshift and scale dependence of the bias.
In Section~\ref{sec:modeling}, we show the anisotropic \HI\ power spectra in redshift space, and compared them with the linear and non-linear clustering models.
Section~\ref{sec:results} is devoted to the interpretations of our results, and then we give a summary in Section~\ref{sec:summary}.
Throughout this paper, we assume the cosmological parameters consistent with the WMAP-9 year result \citep{WMAP9}.

\section{Simulation}
\label{sec:simulation}
In this section, we give a brief summary of two cosmological hydrodynamic simulations that we use in this paper, and describe how we make the mock simulated data targeting the future 21 cm observations.

\subsection{Illustris simulation}
\label{ssec:illustris}
One of the data set we use to evaluate the \HI\ bias is from the  cosmological hydrodynamic simulation Illustris,\footnote{http://www.illustris-project.org/} in which the thermal and dynamical evolution of baryons is solved with a moving-mesh code {\small AREPO}~\citep{Springel+2010, Genel+2014, Vogelsberger+2014}.
We use the data set of Illustris-3 simulation: 
the box-size is $75\,\himpc$ on a side, $2\times455^3$ gaseous cells and dark matter (DM) particles are distributed in the volume. 
The Illustris simulation adopts the WMAP-9 cosmological parameters \citep{WMAP9}: 
$\Omega_m = 0,2726$,  $\Omega_\Lambda = 0.7274$, $\Omega_b = 0.0456$, $\sigma_8 = 0.809$, $n_s = 0.963$, $h = 0.704$. 
The resultant mass resolution is $5.70\times 10^7\,\himsun$ for gas and $2.81\times 10^8\,\himsun$ for DM \citep{Nelson+2015}.

In the Illustris simulation, neutral hydrogen is ionized via photo-ionization and collisional ionization processes. 
The simulation employs a UV background (UVB) model of \cite{FG11} at $z<6$.  
It is noted that the photo-ionization by the UVB mainly contributes to the heating on large scales. 
The gas	temperature, which controls the collisional ionization rates, is determined by the competition between cooling and heating.	
The feedback driven by SNe and AGNs often heats the gas to above $10^5$~K in the vicinities of galaxies, and consequently such regions are highly ionized via collisional ionization. 

The SN feedback model in the Illustris simulation is basically the same as the wind model implemented in SPH-based schemes
\citep[e.g.][]{SH2003,OD2008,Okamoto+2010}, except for a slight modification to account for mesh-based scheme \citep{Vogelsberger+2013,Torrey+2014}. 
As for the AGN feedback, the Illustris simulation adopts a two-state model, in which either radio-mode feedback or quasar-mode feedback is chosen according to the mass accretion rate onto a supermassive black hole (BH). 
When the BH accretion rate is high, quasar-mode feedback is activated so that a fraction (0.1--0.2) of the radiative energy released by the BH accretion is converted to the thermal energy. 
In the contrary case of low BH accretion rates, a jet launched from the BH mechanically affects the surrounding medium. 
In addition to these thermal and mechanical AGN feedback, the photo-heating and photo-ionization by the radiation from AGNs are also considered \citep{Vogelsberger+2013,Torrey+2014}. 

Although the feedback efficiencies are adjusted so as to reproduce the stellar mass function at the present-day and the cosmic star formation history \citep{Genel+2014}, the AGN feedback in the Illustris simulation is known to be too strong, which resulted in the overheated IGM \citep{Viel+:2017} and the hot gas was transported too far from galaxies \citep{Haider+2016}. 

\begin{figure*}
  \begin{tabular}{ccc}
    \includegraphics[width=0.3\linewidth]{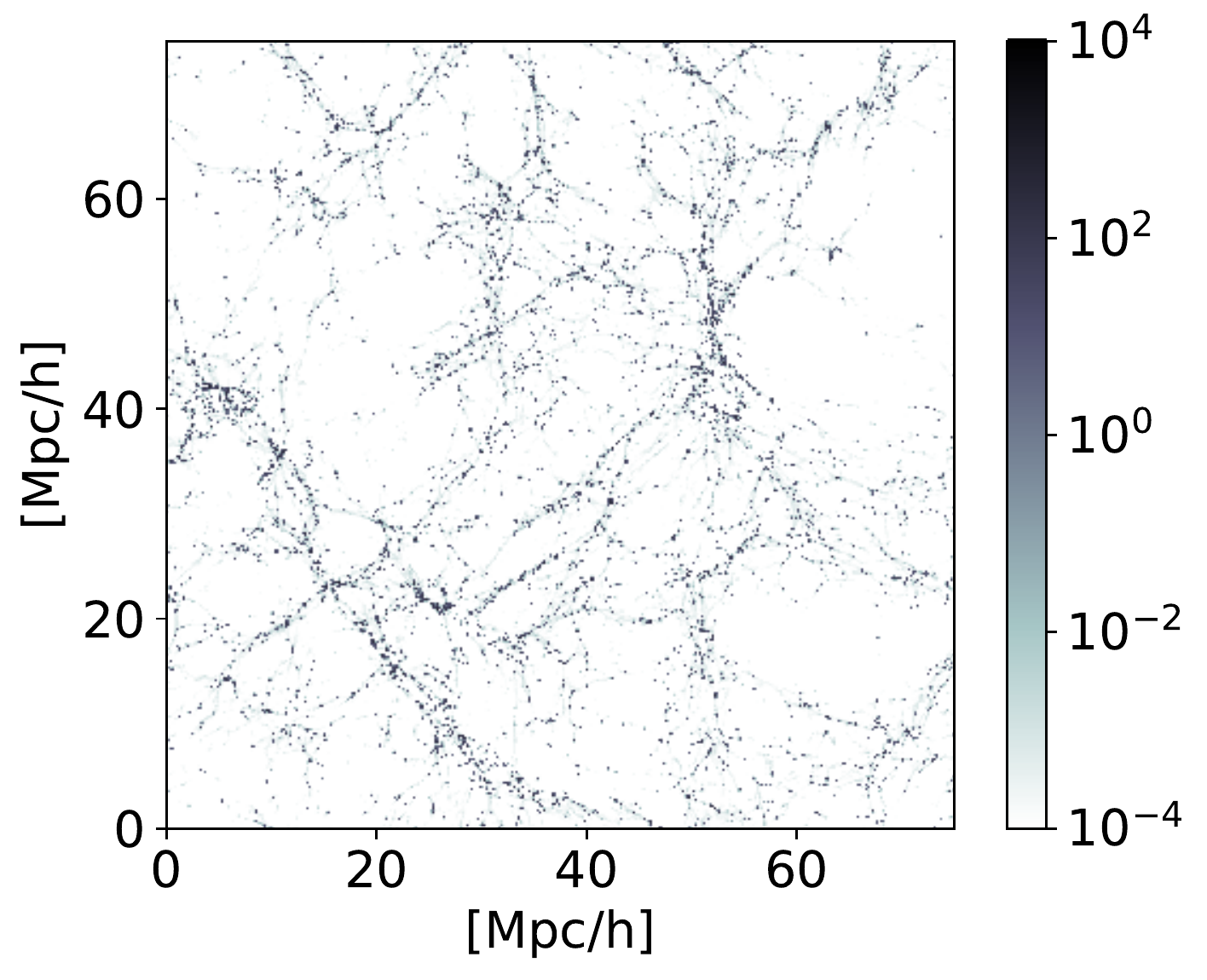} &
    \includegraphics[width=0.3\linewidth]{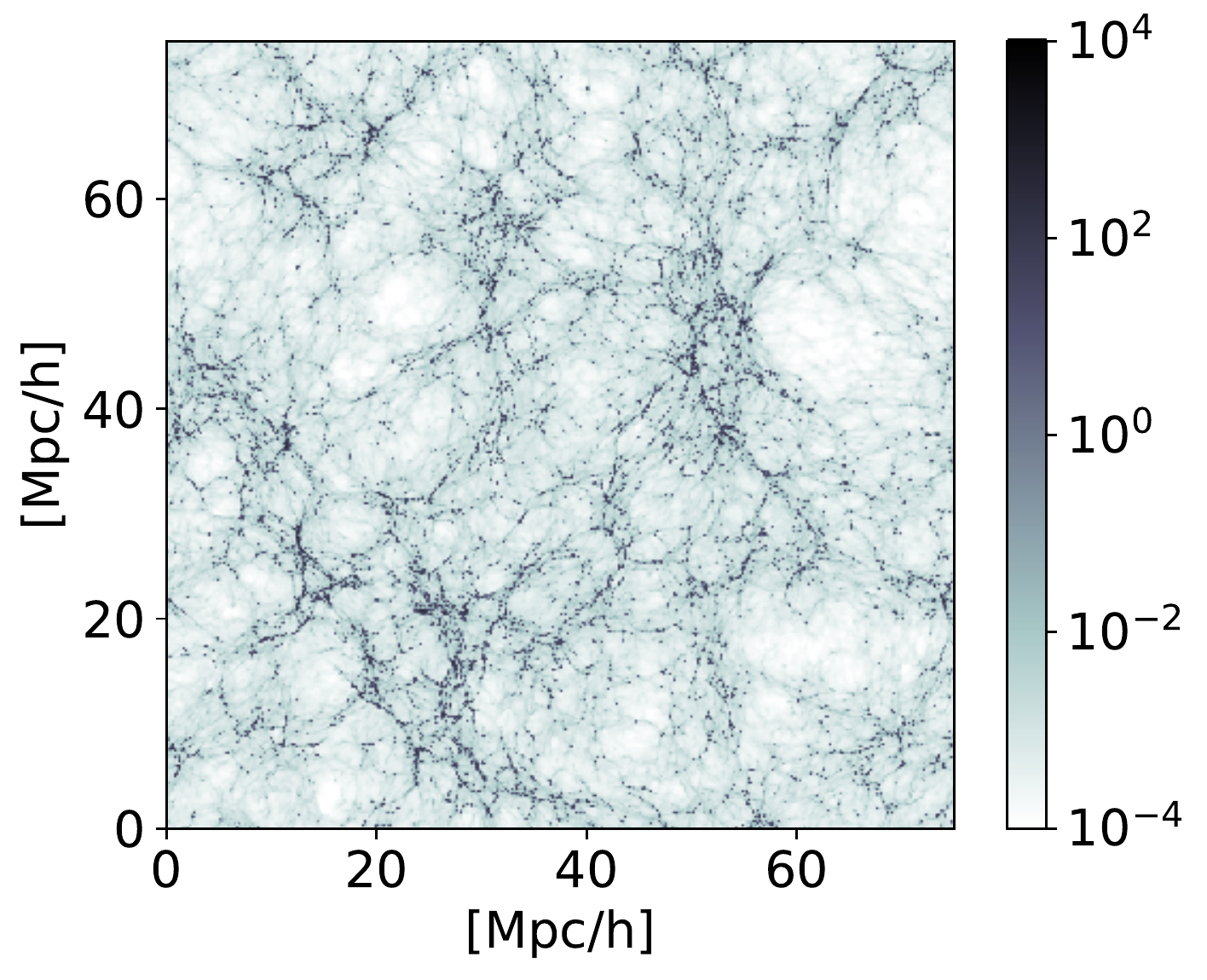} &    
    \includegraphics[width=0.3\linewidth]{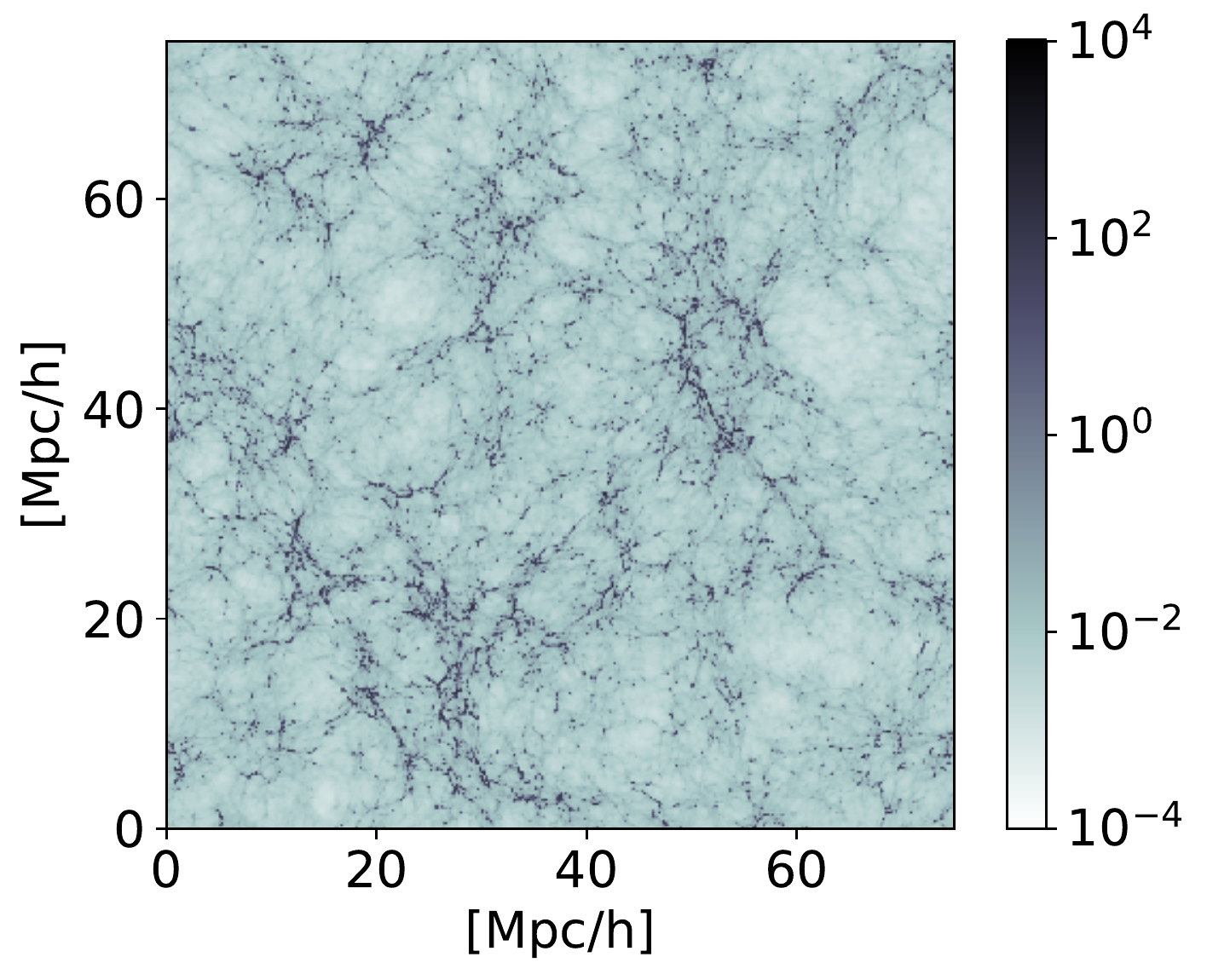} \\
    \includegraphics[width=0.3\linewidth]{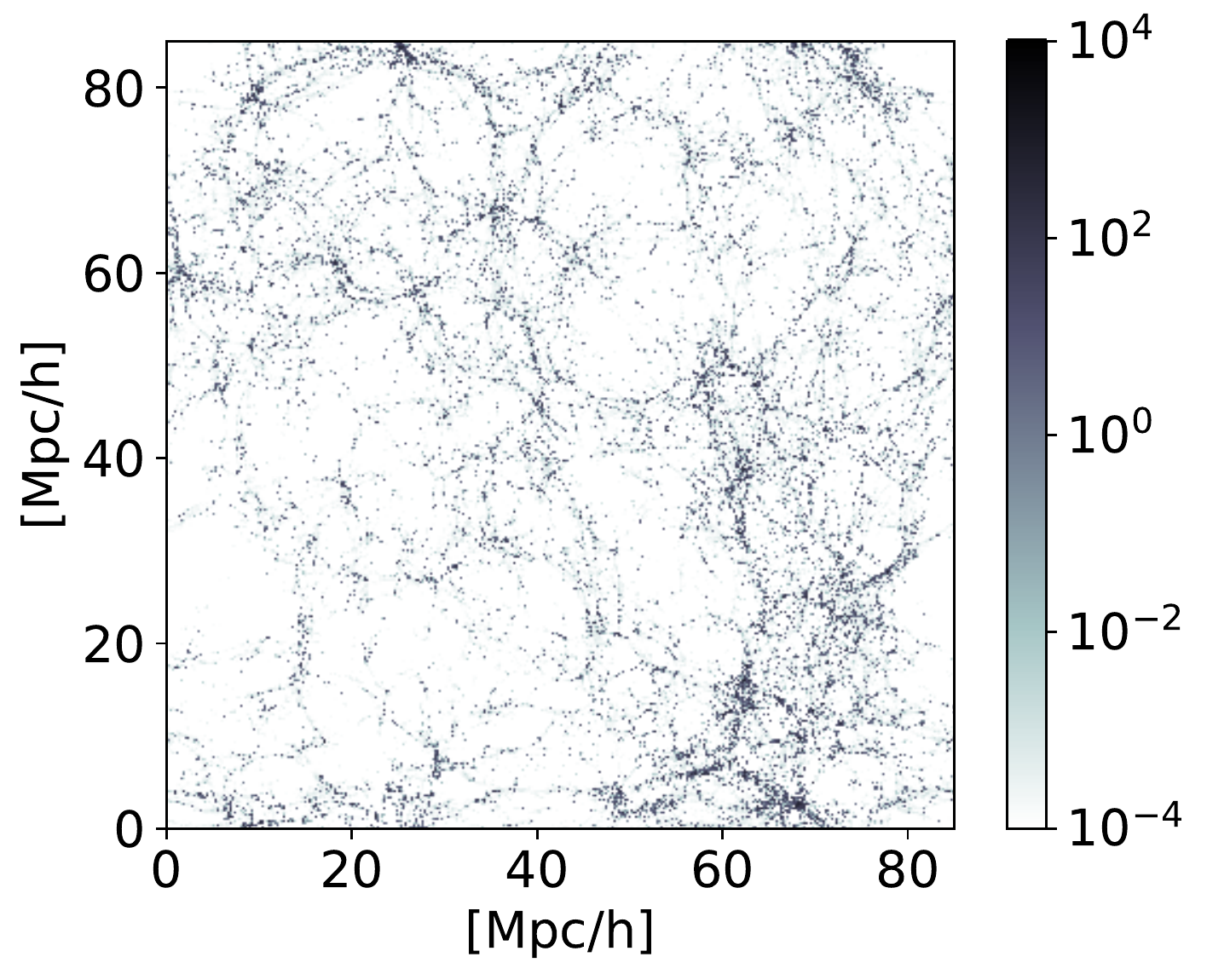} &    
    \includegraphics[width=0.3\linewidth]{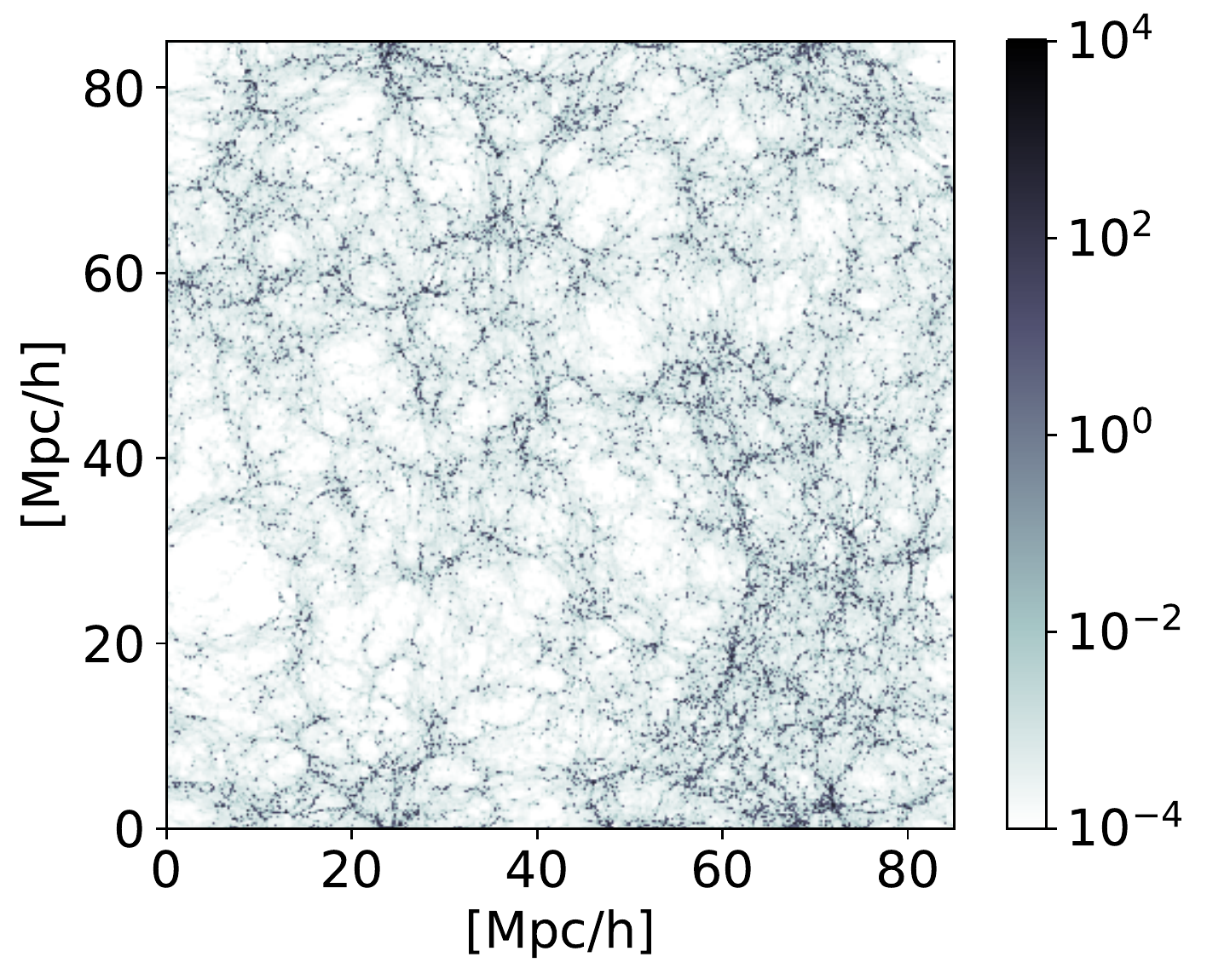} &
    \includegraphics[width=0.3\linewidth]{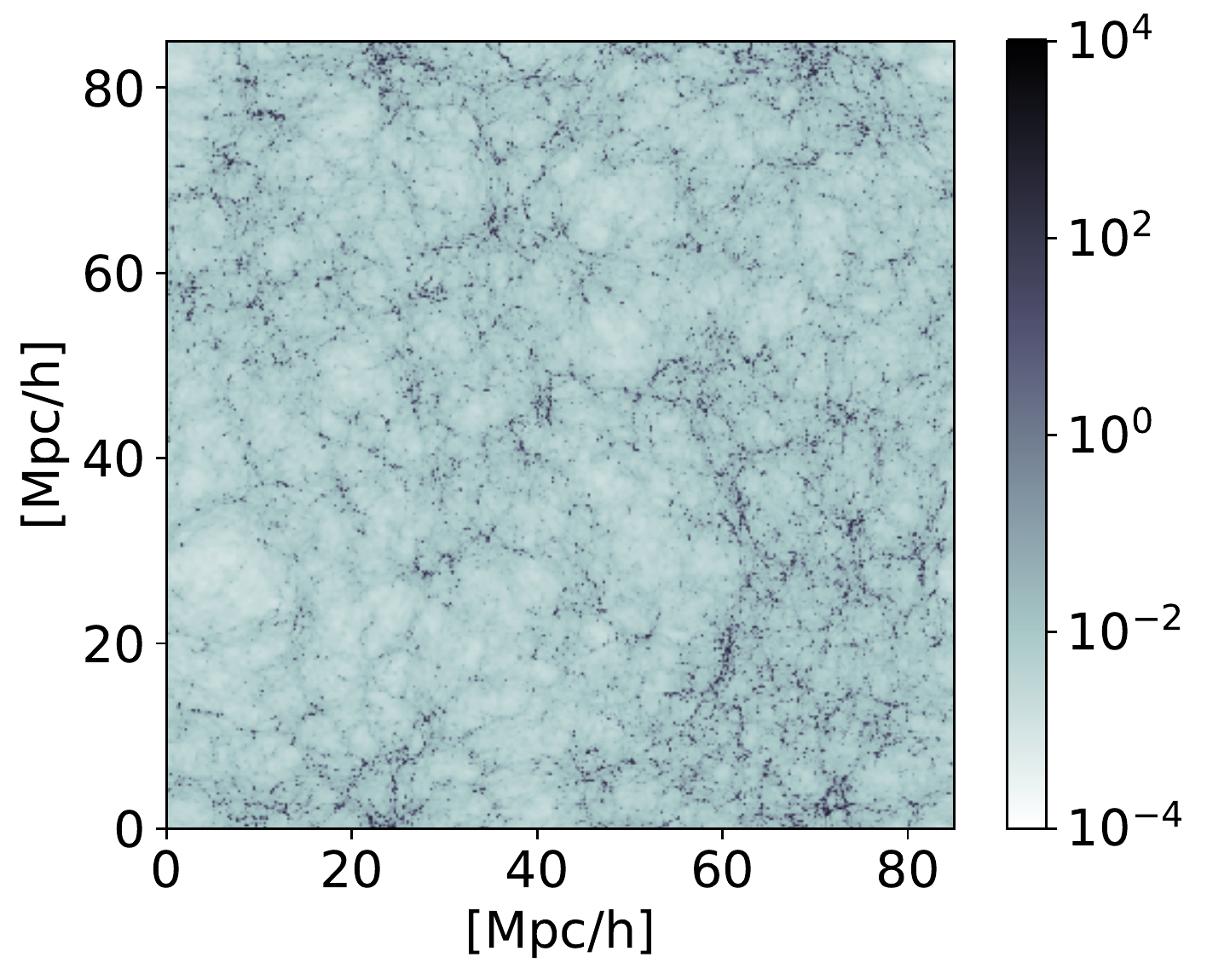} \\
  \end{tabular}
  \caption{
  	\HI\ gas distribution at redshift $z=1, 3$ and $5$ from left to right panels, for Illustris simulation (upper panels) and Osaka simulation (lower panels). Color gradient represents the \HI\ gas overdensity $1+\delta=\rho_{\HI}/\bar{\rho}_{\HI}$ in logarithmic units, where $\bar{\rho}_{\HI}$ is the mean \HI\ density at each redshift.  
  \label{fig:maps}}
\end{figure*}

\subsection{Osaka simulation}
\label{ssec:osaka}

The Osaka simulation uses a modified version of N-body/SPH code {\small GADGET-3} \citep[originally described by][]{Springel2005}.  Our code includes treatment for star formation and supernova feedback (but no AGN feedback), and the details can be found in \citet{Aoyama+2017} and Shimizu et al. (2018, submitted). 
The uniform UV background radiation of \citet{HM12} is used, and the cooling is solved using the Grackle chemistry and cooling library \citep{Smith+2016}.\footnote{https://grackle.readthedocs.org/}

The simulation used in this paper has a box-size of comoving $85\,\himpc$, and the initial particle number is $2\times 512^3$ for gas and dark matter.  The number of gas particles decreases slowly as some of them are converted into star particles in high-density regions. 
We use the same  WMAP-9 cosmological parameters as the Illustris simulation for a fair comparison. 
The particle masses are $5.79\times 10^7\,\himsun$ and $2.88\times 10^8\,\himsun$ for gas and dark matter particles, which are roughly equivalent to those in the Illustris simulation. 

Here we briefly review the prescription for star formation and SN feedback. The star formation is allowed when the local gas density exceeds the threshold density of $n_{\rm SF,th}=0.1$\,cm$^{-3}$, and two star particles can be created from each gas particle. The star particles are created statistically such that the time-averaged star formation rate (SFR) will recover the following rate: 
\begin{equation}
\dot{\rho}_{\star} = \epsilon_\star \frac{\rho_{\rm gas}}{t_{\rm ff}},
\end{equation}
where we take $\epsilon_\star=0.05$ and $t_{\rm ff} = \sqrt{3\pi/(32G\rho_{\rm gas})}$ is the local free-fall time. 
In this work, we do not consider the self-shielding correction, as we are not concerned too much about the small scales ($< 100$\,kpc) when we compute the power spectrum and the bias parameter. 
Once the star particle is created, we compute the total SN energy that will be deposited based on the Chabrier IMF,  
and assign 30\% (70\%) of that energy to gas particles within the shock radius as the kinetic (thermal) energy. 
The metals are also distributed in the same manner as the SN energy.

\subsection{Mock data for future observations}
\label{ssec:ska}
In order to compare the two simulations on the same ground, 
we first define a $512^3$ grid in the simulation box to recompute the
{\sc Hi} density field, where each grid-cell size is 146.5\,$h^{-1}$kpc for Illustris and 166.0\,$h^{-1}$kpc for Osaka simulation, respectively. These grid scales are much finer than the scale of our interest, therefore it will not affect our conclusion on the large-scale bias. 
The density field is represented by the SPH particles (or the fluid element in the Voronoi cells for the Illustris simulation) which have corresponding density and smoothing scale.  For the Osaka simulation, we use the density and the smoothing scale of each SPH particle, and recompute the {\HI} density field on our grid using the following cubic spline kernel \citep{SPHKernel:1985}:
\begin{equation}
  \label{eq:SPHKernel}
  W^{\rm SPH}(r;h) = A\left\{
    \begin{array}{ll}
      \displaystyle{1-\frac32 \left(\frac{r}{h/2}\right)^2 + \frac34
      \left(\frac{r}{h/2}\right)^3} & 0 < r < \frac{h}2 \\
      \displaystyle{\frac14 \left(2-\frac{r}{h/2}\right)^3} & \frac{h}2 < r < h \\
      0 & h < r \\
      \end{array}
    \right.,
\end{equation}
where $h$ is the smoothing length for each particle, and $r$ is the distance between the particle and the grid. 
The amplitude $A$ is determined such that for every particle, the sum of $W^{\rm SPH}$ over all grid becomes unity.

For the Illustris simulation, we define the smoothing length of each fluid element as 
\begin{equation}
	\label{eq:sml_vol}
		  \RA{\frac{h}2=\left(\frac{3V}{4\pi}\right)^{\frac13},}  
\end{equation}
where $V$ is the volume of each Voronoi cell. 
We checked that the above methods give similar PDFs of \HI\ density in both simulations.  
Using this smoothing length and Eq.~(\ref{eq:SPHKernel}), we compute the \HI\ density in each grid-cell similarly to the Osaka simulation.  

The {\HI} gas density in each grid is then calculated by summing over all contributing particles,
\begin{equation}
	\label{eq:SPHdensity}
  \rho_{\HI}^{\rm SPH}(\bs{x}_i)
  = 
  \sum_j W^{\rm SPH}(|\bs{x}_i-\bs{x}_j|; h_j) \rho_j n_{{\HI}, j},
\end{equation}
where $\rho_j$ and $n_{{\HI},j}$ denote the total gas density 
and neutral hydrogen fraction assigned to the $j$-th particle
located at $\bs{x}_j$ in the simulation box. 
For the particle which has too small smoothing scale at high-density region, the mass is deposited only to the local cell. 
Figure~\ref{fig:maps} shows the {\HI} density contrast $1+\delta=\rho/\bar{\rho}$ in logarithmic scales from the two simulations, where $\bar{\rho}$ is the mean density at each redshift. 
The obtained density contrast of \HI\ defined on the regular grid is Fourier transformed using the python module \texttt{numpy.fftn}, and we obtain the three dimensional density field in Fourier space. 
We note that the different scheme to define the gas density with adaptive mesh refinement can be found in \cite{Behrens+:2018}.

Here we further consider the angular and frequency resolutions for the \textit{Square Kilometre Array} (SKA)-like observation for both interferometer and single-dish observations.  
For other future observations, we can redefine the  resolution of the grid according to the specification of each observation.
For the interferometer mode, the angular resolution is quite high, and we assume here that the {\HI} cloud is identified at the resolution of 3', which corresponds to the comoving scale of $2\,\himpc$ at $z=1$ and $4.8\,\himpc$ at $z=5$, respectively. However, the field of view of interferometer is small and not optimal for a large sky coverage. On the other side, the single dish observation has a wide field of view, but the angular resolution is relatively coarse.
For single-dish observations, given that the size of dish is fixed, we assume that the angular resolution is proportional to the
observed wavelength as $\theta = \lambda_{\rm obs} / D$, where $\lambda_{\rm obs}$ is the observed wavelength of 21\,cm line and 
$D$ is the diameter of the dish, $D=15$ meters for SKA. This corresponds to the angular size of	
$48.1(1+z_{\HI})$ arcmin, which is comoving $65\,\himpc$ at $z=1$ and $460\,\himpc$ at $z=5$, respectively.
The spatial resolution along the line of sight (LoS) is determined by the
frequency resolution of the observation. We assume 50 kHz for all
frequency channels which is much finer than our initial choice of the
grid.  Thus we do not redefine the radial resolution.

\section{Scale Dependent Bias in Real Space}
\label{sec:pspec}
For the unbiased measurement of the dark energy parameters through the location of BAO peaks and troughs, 
it is important to understand the scale dependence of {\HI} bias on large scales where the matter clustering is in quasi-linear regime.  
\rev{Moreover if we can precisely model the {\HI} bias, the full shape of power spectrum provide us with more cosmological information than the BAO alone.}
In this section, we first measure the {\HI} power spectrum 
and evaluate the significance of scale dependence of {\HI} clustering without considering the redshift space 
distortion. The effect of RSD will be considered in Section~\ref{sec:modeling}.

\subsection{Measurement of Power Spectrum and \HI\ bias}
\label{ssec:pspec1d}
\begin{figure*}
	\includegraphics[width=\linewidth]{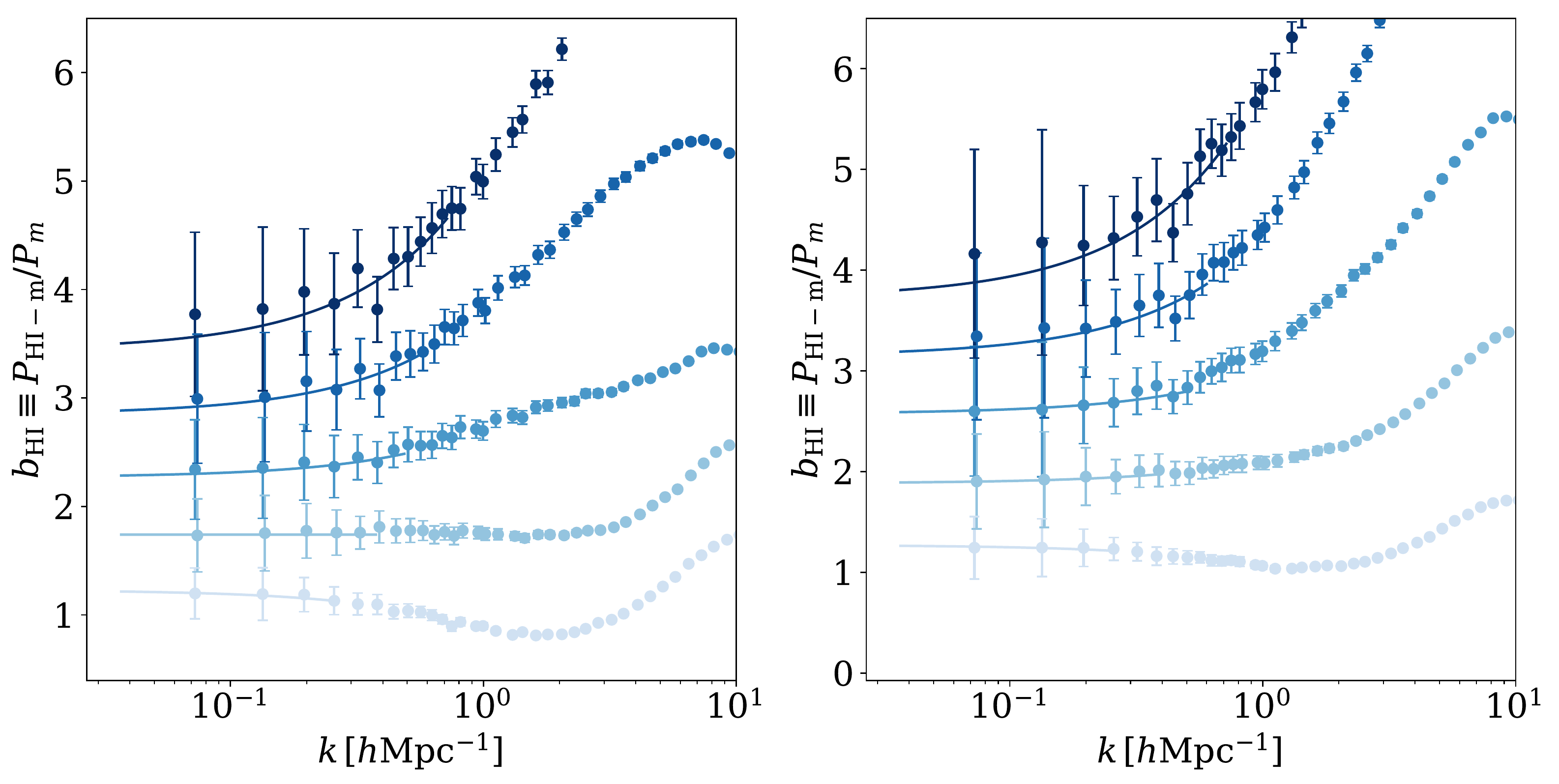}
    \caption{
    \HI\ bias defined by the cross correlation (Eq.~\ref{eq:bias_cross}) measured from the Illustris-3 simulation (\textit{left}) and the Osaka simulation (\textit{right}).
    In both panels, the lines correspond to $z=5, 4, 3, 2,$ and  $1$ from top to bottom, respectively. The solid curves are the best-fitting models defined by Eq.~(\ref{eq:bias_polynomial}). 
    \rev{{The error bars are the standard deviation of $b(k)$ within each bin divided by the number of modes available, $\sigma/\sqrt{N_k}$, which means the standard deviation of the estimated mean. The positions of data points and error bars are slightly shifted for the same bin in order to avoid complete overlap of error bars.}}
	    \label{fig:bias_kmax}}
\end{figure*}
As we only work with the simulation data, it is useful to work in the Fourier space. 
Thus we define the {\HI} bias using the power spectrum.
In real space, where we do not consider the RSD effect, only the absolute value of wavenumber is considered from $k_{\rm min}=2\pi/L_{\rm box}$ to $k_{\rm max}=\pi N_{\rm grid}/L_{\rm box}$. 
The power spectrum is then calculated as the algebraic mean among the absolute value of $k$, 
\begin{equation}
	\label{eq:def_powerspectrum}
	P_{\rm XY}(k_i) = \frac{1}{N_k} \sum_{j, k_j \in k_i}^{N_k} \Re{[\delta_X(k_j) \delta_Y^*(k_j)]},
\end{equation}
where ${\rm X,Y}$ denote either total matter or {\HI} density fluctuation.
The \HI\ bias can then be defined using the measured power spectra as
\begin{equation}
	\label{eq:bias_cross}
  \hat{b}_{\HI}^{\rm cross}(k) 
  \equiv 
  \frac{P_{\HI,\, {\rm m}}(k)}{P_{\rm m}(k)},
\end{equation}
where $P_{\HI,\, {\rm m}}$ and $P_{\rm m}$ are the power spectra of \HI--matter cross correlation and matter auto correlation, respectively. Note that the matter is a sum of dark matter and gas components in the simulation.
Another way of defining the \HI\ bias would be
\begin{equation}
	\label{eq:bias_auto}
	\hat{b}_{\HI}^{\rm auto}(k) \equiv 
   \sqrt{\frac{P_{\HI}(k)-S}{P_{\rm m}(k)}},
\end{equation}
where $P_{\HI}$ is the auto power spectrum of \HI\ gas, and $S$ is the corresponding shot-noise.
The two different definitions of bias are identical on large scales where the density fluctuation is described by the linear perturbation theory and the effect of shot-noise is not dominant. 
In the language of higher order perturbation theory, on the quasi-linear scales, the mode coupling between different wavenumber modes enters differently for cross-correlation and auto-correlation for the biased tracers. This makes the different definitions of bias behave differently. 
In this paper, we take the former definition, because 
the cross correlation is totally free from the shot-noise effect. 
Hereafter we write $\hat{b}_{\HI}\equiv \hat{b}_{\HI}^{\rm cross}$ for simplicity. 

Figure~\ref{fig:bias_kmax} shows the measured \HI\ bias using the Illustris and the Osaka simulations. 
Due to the limited box-size of the simulations, we cannot go to the larger scale of $k<0.1 \hmpci$; however, the simulations clearly show that the \HI\ bias seems to converge to constant values on large scale. 
The constant convergent value depends largely on the redshift, and we see that the bias is higher at higher redshifts. 
We will discuss in the next section in more detail, but the Illustris simulation has systematically lower bias values compared to those in the Osaka simulation at all redshift ranges. 

\rev{{In passing, we also compared the {\HI} bias in both Illustris-1 (higher resolution) and Illustris-3 simulation, and find that the Illustris-3 gives higher $b_{\HI}$ due to the effect of mass resolution, but the discrepancy is less than 10\% which is well within the 1-$\sigma$ statistical error.  For the reason that we explained in \S~\ref{sec:introduction}, we use Illustris-3 for our fiducial comparison. 
}}

At $1<z<3$, since the Universe is almost perfectly ionized \citep{Fan+2006,Becker+2015} and the amount of the neutral hydrogen in the IGM is negligibly small, the majority of {\HI} gas is confined in the high-density regions such as inside the galactic halos \citep[e.g.][]{Nagamine2004a,Nagamine2004b}.
Nevertheless, compared to the bias of QSOs, the {\HI} bias is still lower \citep[e.g.][]{Laurent+:2017}. This implies that the {\HI} gas is more broadly distributed than QSOs, or in other words, the QSOs reside only in extremely high-density environment.

\subsection{Scale dependence of \HI\ bias}
\label{ssec:model-bias}
It is known that the scale dependence of bias may shift the scale of
BAO peak, and therefore, an accurate modeling is crucial for precisely constraining the cosmological parameters.
To find the signature of scale dependence of \HI\ bias, 
here we introduce the linear function of $k$ as 
\begin{equation}
	\label{eq:bias_polynomial}
  b_{\HI}(k) = b_0 + b_1 k,
\end{equation}
where $b_0$ and $b_1$ are free parameters. 
\rev{{
Here we introduce the $k$-dependence of bias simply as `$b_1 k$' to discriminate between the constant and scale-dependent bias.  It is known that the $k$-dependent term is induced by the relative velocity of baryon and dark matter \citep{Fabian+2016}.
}}
We fit the measured \HI\ bias from simulations with Eq.\,(\ref{eq:bias_polynomial}) to quantify the scale dependence. 
We find the best-fitting parameters by a usual Metropolitan-Hastings MC method with the likelihood 
\begin{equation}
	\label{eq:chisq_bias}
   {\mathcal L} = \exp\left[ - \frac12 \sum_i^{k_i<k_{\rm max}} 
   \frac{\left( \hat{b}_{{\HI}, i}-b_{\HI}(k_i) \right)^2}{\sigma_i^2} \right],
\end{equation}
where $k_{\rm max}$ is the maximum wavenumber for the fitting. 
We first choose $k_{\rm max}$ so that the fluctuation of dark matter is not too large,
\begin{equation}
\label{eq:k_max}
	\frac{k_{\rm max}^2}{6\pi^2} \int^{k_{\rm max}}_0 \! {\rm d}k P^{\rm lin}(k,z) = C,
\end{equation}
with $C=0.7$. This criterion is empirically derived, such that the dark matter power spectrum of $N$-body simulation and prediction of higher-order perturbation theory agrees within 1\% \citep{Nishimichi+:2009,Taruya+:2009}. This is slightly conservative for our study, but it reasonably suggests the scale where structure becomes quasi-linear and the \HI\ bias supposedly having a scale dependence.

\begin{figure*}
\begin{tabular}{cc}
\includegraphics[width=0.45\linewidth]{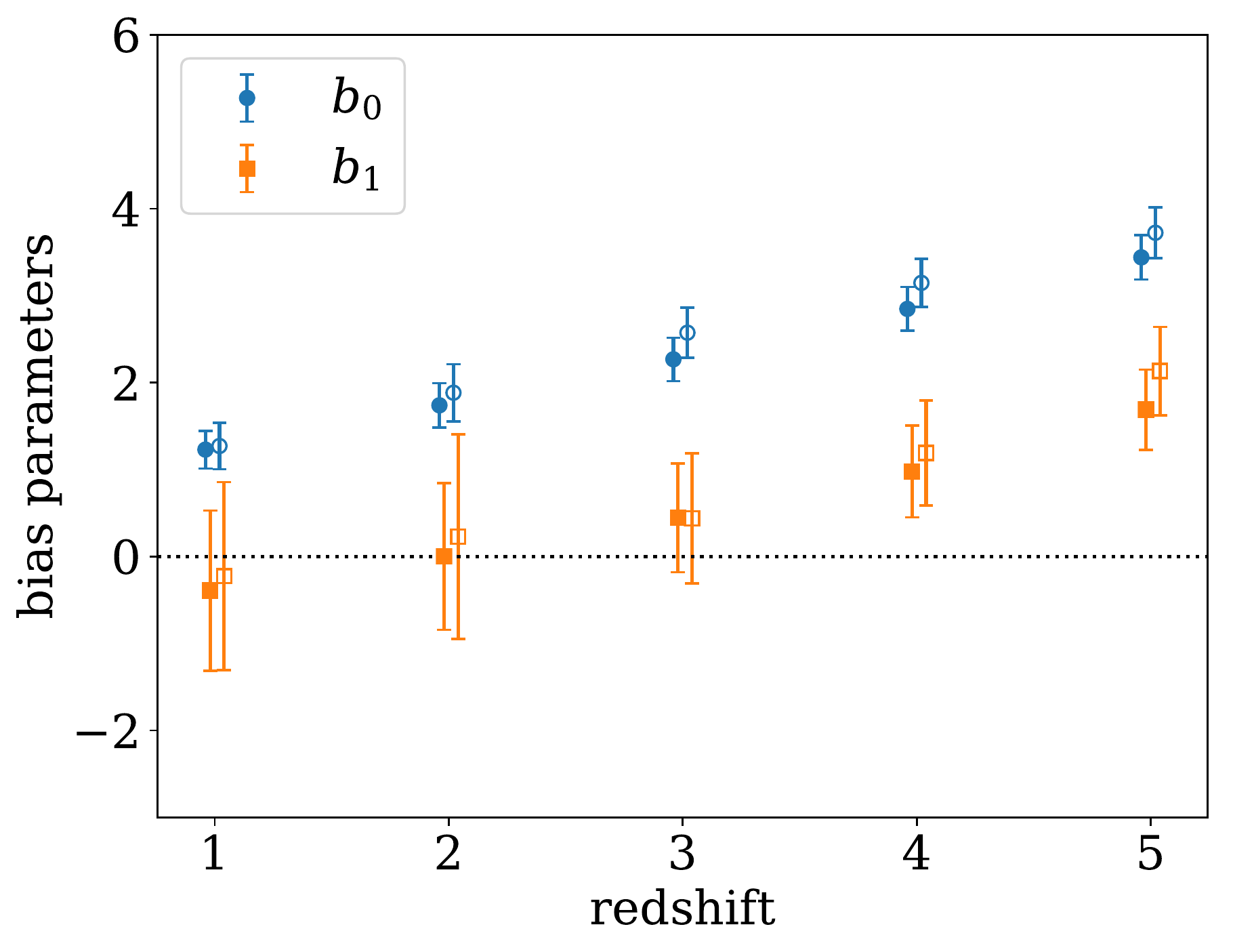}&
\includegraphics[width=0.45\linewidth]{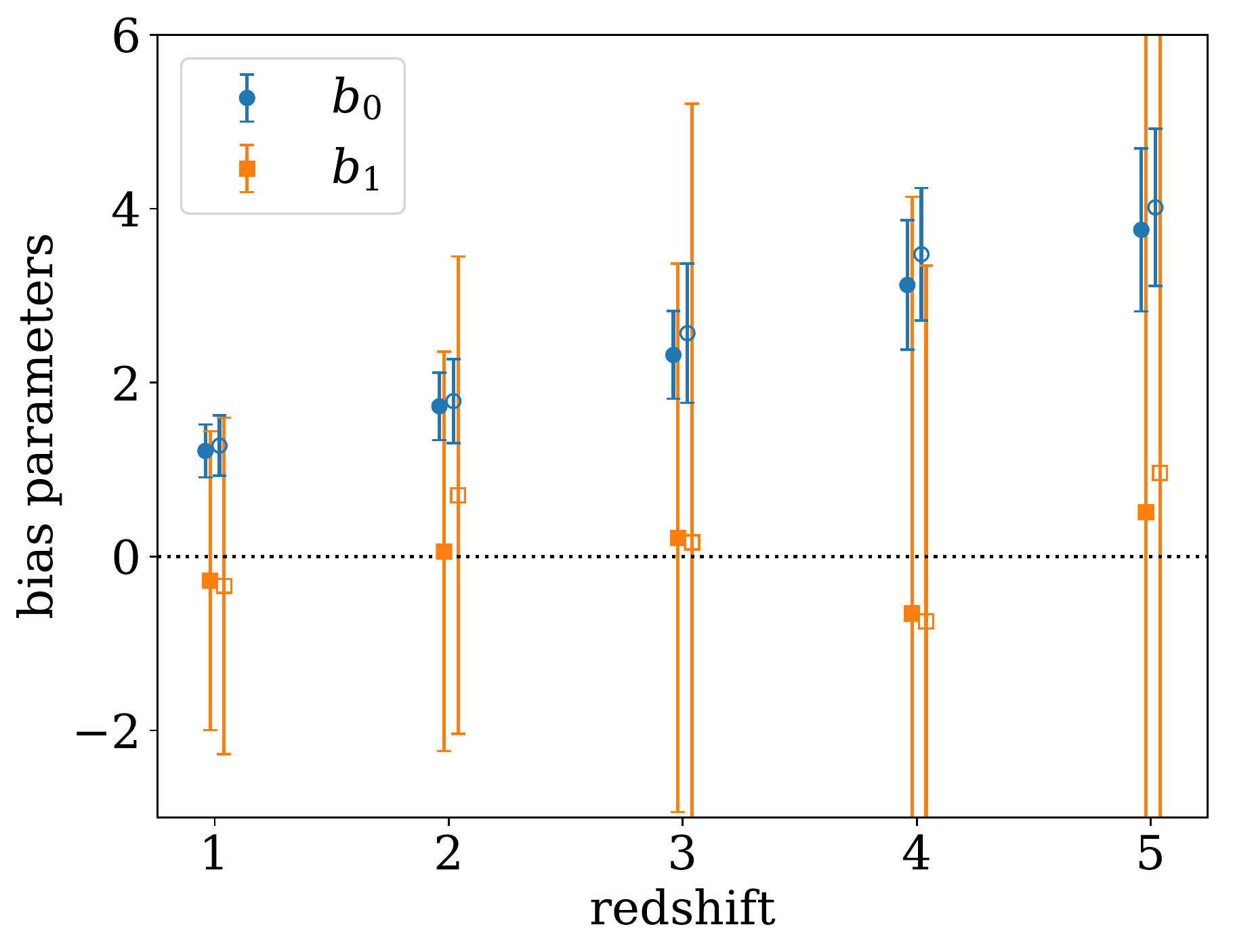}
\end{tabular}
\caption{
	{\it Left panel}: Best-fitting \HI\ bias parameters for Illustris (filled) and Osaka (open) simulations.
    The value of $k_{\rm max}$ used for the fitting is given by the limit defined in Eq.~(\ref{eq:k_max}).
    The bias parameters are consistent with each other in the two simulations, and we clearly find the scale dependence only at $z>3$ while it is consistent with being constant (i.e. $b_1 =0$) at $z<3$. The data points are slightly shifted horizontally for illustrative purposes. {\it Right panel}: Same as the left panel but for $k_{\rm max}=0.25\,\hmpci$. The bias is consistent with being constant at all redshift ranges.
\label{fig:bias_kdept}}
\end{figure*}

Figure~\ref{fig:bias_kdept} shows the best-fitting parameters for the \HI\ bias. 
\rev{{Since we do not apply any priors on the parameters, Eq.\,(\ref{eq:chisq_bias}) gives a posterior distribution for each parameters. The statistical errors in Fig.\,\ref{fig:bias_kdept} is properly computed from this full posterior such that 68\% probability is included within the range of the errorbars.} }
As one can see, both the constant bias and scale-dependent components are consistent between two different simulations. We find that the scale dependence of \HI\ bias at $z<3$ is insignificant and consistent with constant bias. On the other hand, at $z>3$, we find a significant scale dependence of the bias. 

\rev{{
In summary, we find that from Fig.\,\ref{fig:bias_kmax}, the \HI\ bias on small scales behaves quite differently between Osaka and Illustris simulations because of different prescriptions for star formation and AGN/SN feedback. On the other hand, from Fig.\,\ref{fig:bias_kdept}, we find that the bias on large scales is consistent with each other within statistical errors, which implies that the details of astrophysics (e.g. star formation and AGN/SN feedback) does not affect the large-scale clustering amplitude very much. However, we note that the errors derived from a single realization of simulation is still large, and a larger number of realizations are needed to see if the difference is statistically significant or not.
}}
We will further discuss this point in Section~\ref{ssec:astro}.

\section{\HI\ power spectrum in redshift space}
\label{sec:modeling}
\subsection{Anisotropic Power Spectrum in Redshift Space}
\label{ssec:pspec2d}
When we measure the cosmological distance to an object, we use the redshift which can be decomposed into cosmological recession velocity and local peculiar velocity of the object. In a limit of Cartesian coordinate, the position of the object can be written as 
\begin{equation}
	\label{eq:peculiar}
   (s_1, \,s_2, \,s_3)=\left(\chi_1,\,\chi_2,\,\chi_3\!+\!\frac{v_3}{aH}\right),
\end{equation}
where $s_i$ and $\chi_i$ $(i=1,2,3)$ are the comoving distance in redshift space and real space, respectively. The coordinate along the line-of-sight (LoS) is $i=3$. The modification to the distance due to the peculiar velocity affects only the separation of two objects along the LoS, which makes the two dimensional correlation function or power spectrum in redshift space distorted \citep{Kaiser+1987} in addition to the geometrical distortion \citep{Matsubara:2004}.

The Osaka simulation uses the SPH method for hydrodynamics, and the gas density and neutral hydrogen fraction is represented by  each gas particle.  Thus we can map the real space density distribution to that in the redshift space simply by moving the gas particle along the LoS direction by $v_3/aH$. The grid based density can then be computed in the exactly same manner described in Section~\ref{ssec:ska}.

\begin{figure*}
  \begin{tabular}{ccc}
    \includegraphics[width=0.3\linewidth]{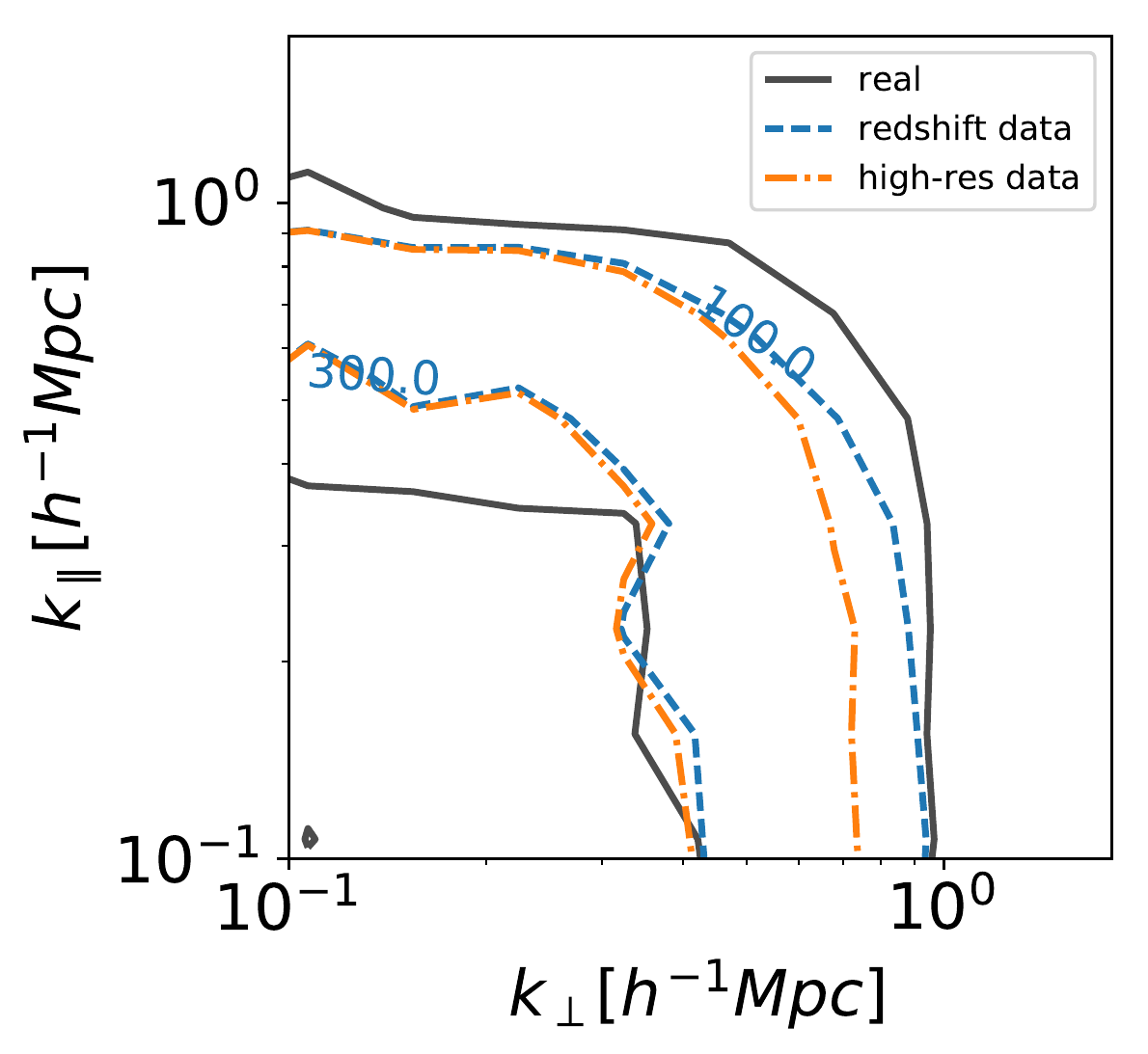}&
    \includegraphics[width=0.3\linewidth]{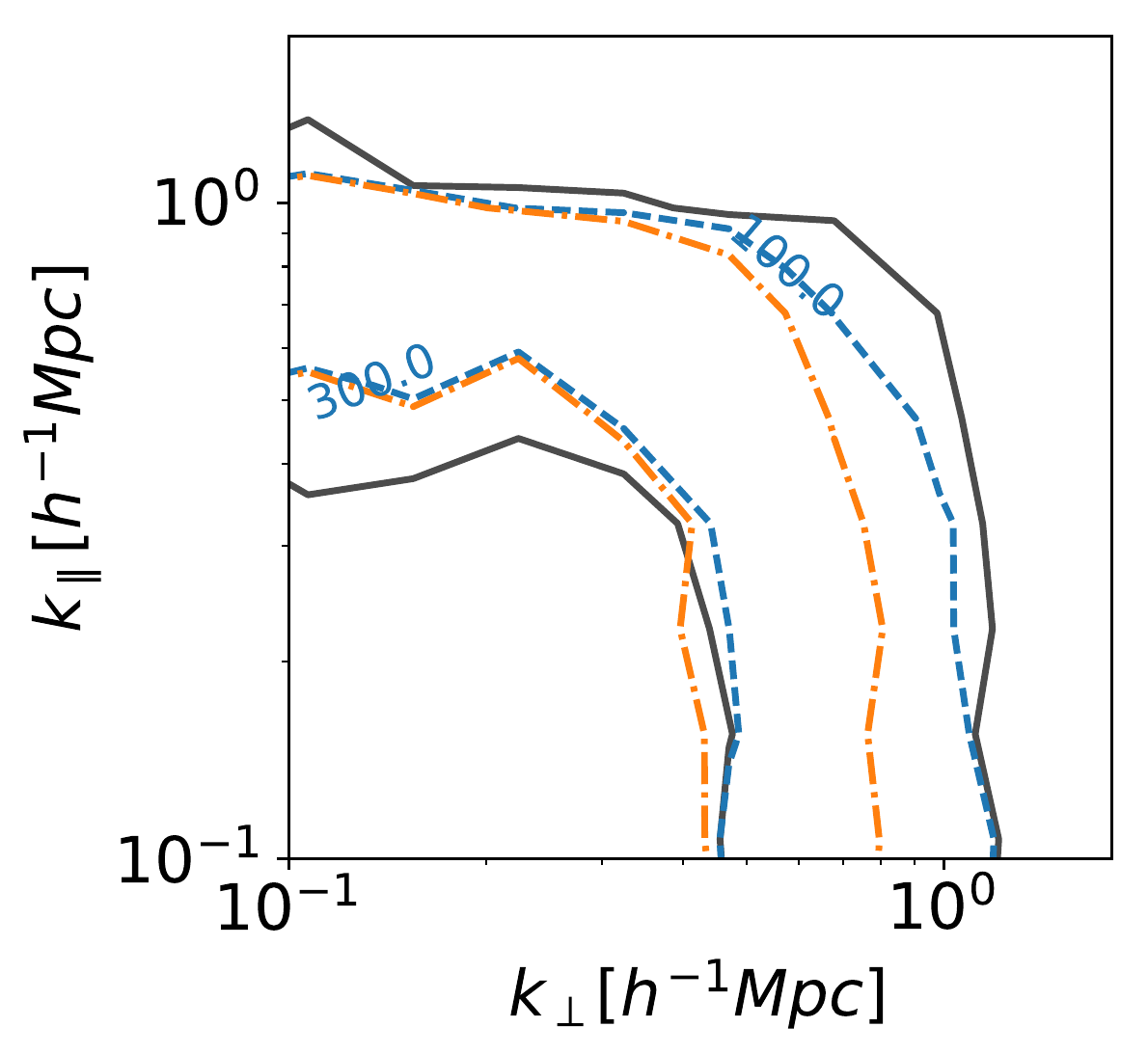}&
    \includegraphics[width=0.3\linewidth]{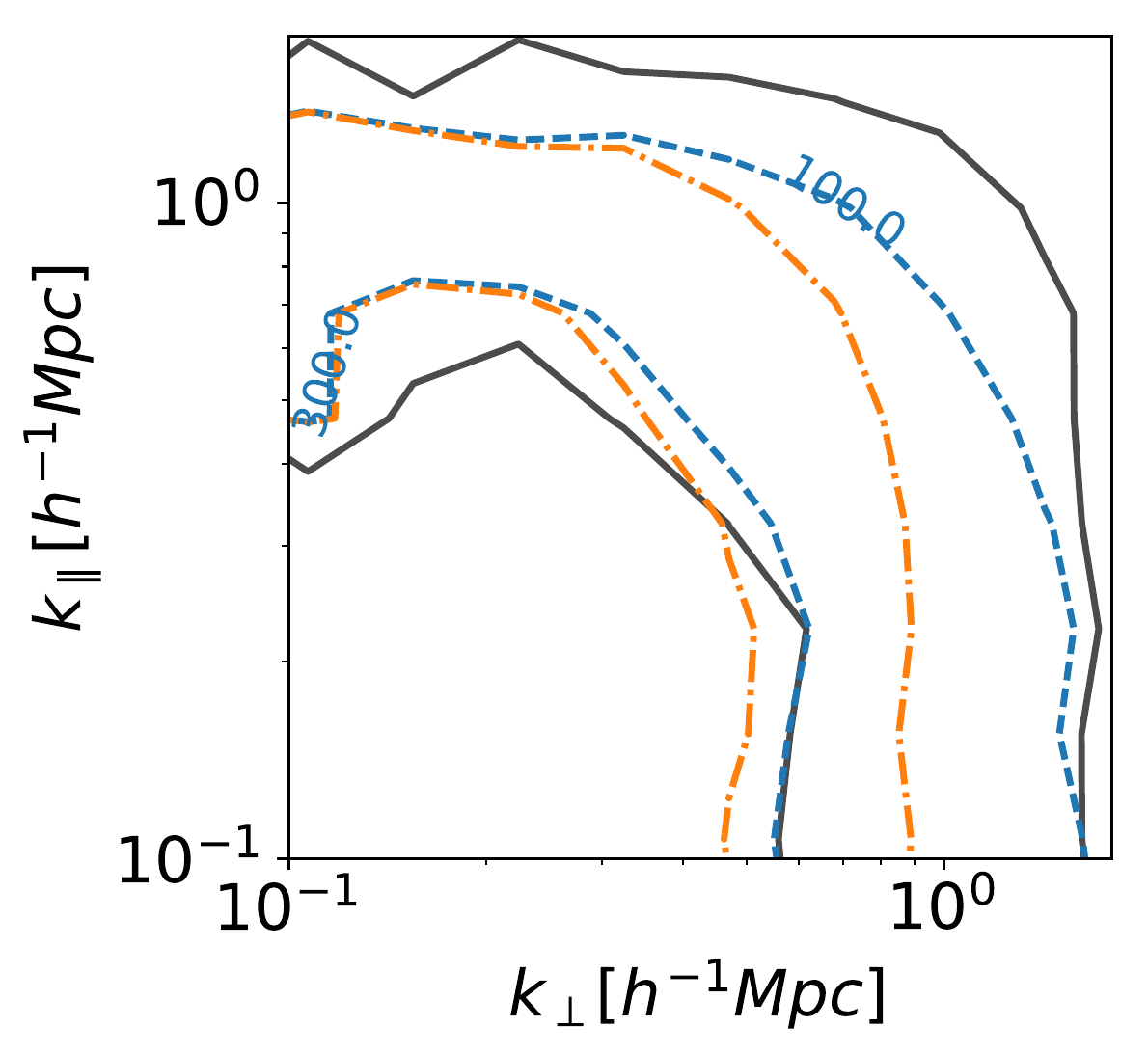}\\
    \includegraphics[width=0.3\linewidth]{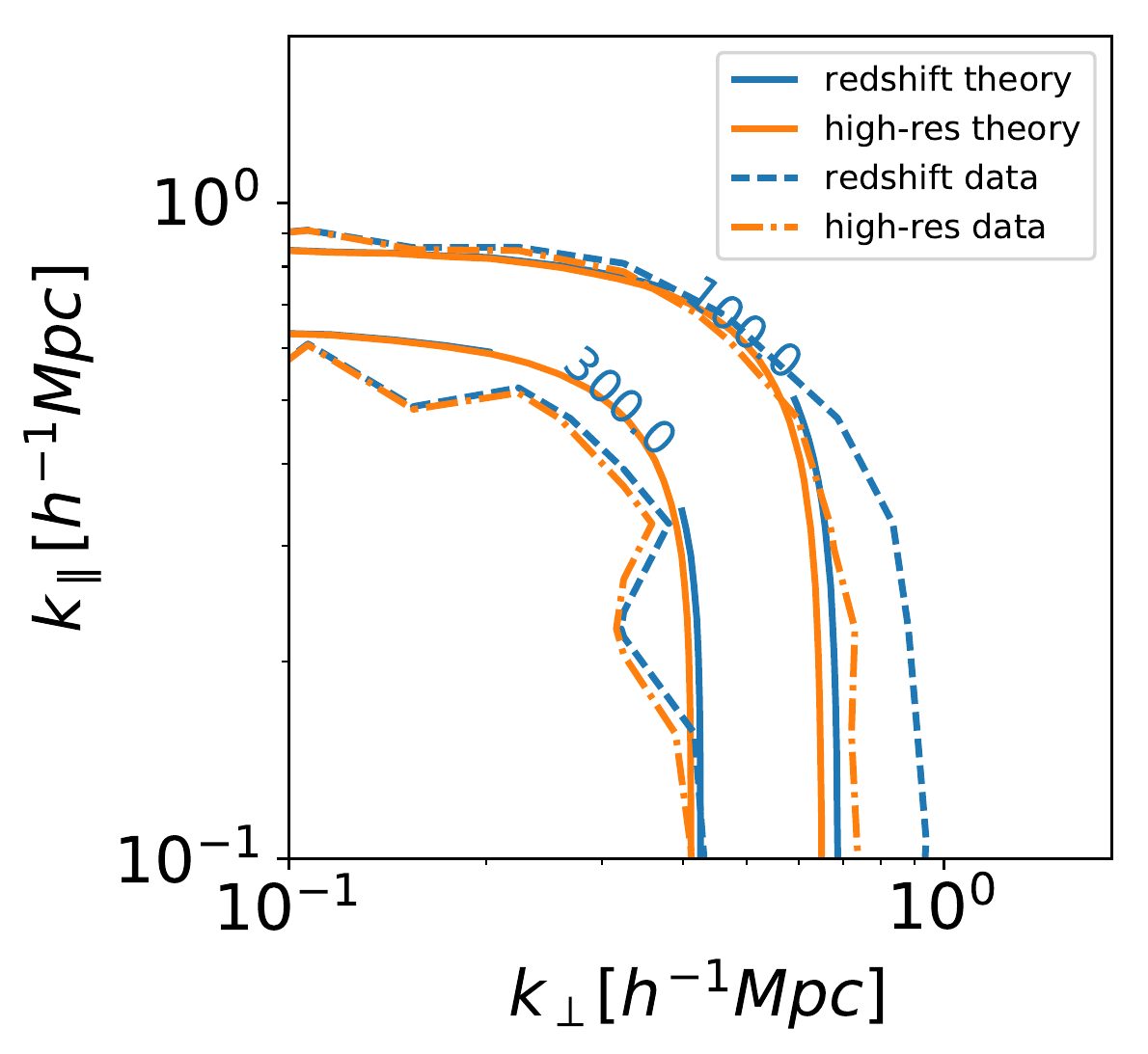}&
    \includegraphics[width=0.3\linewidth]{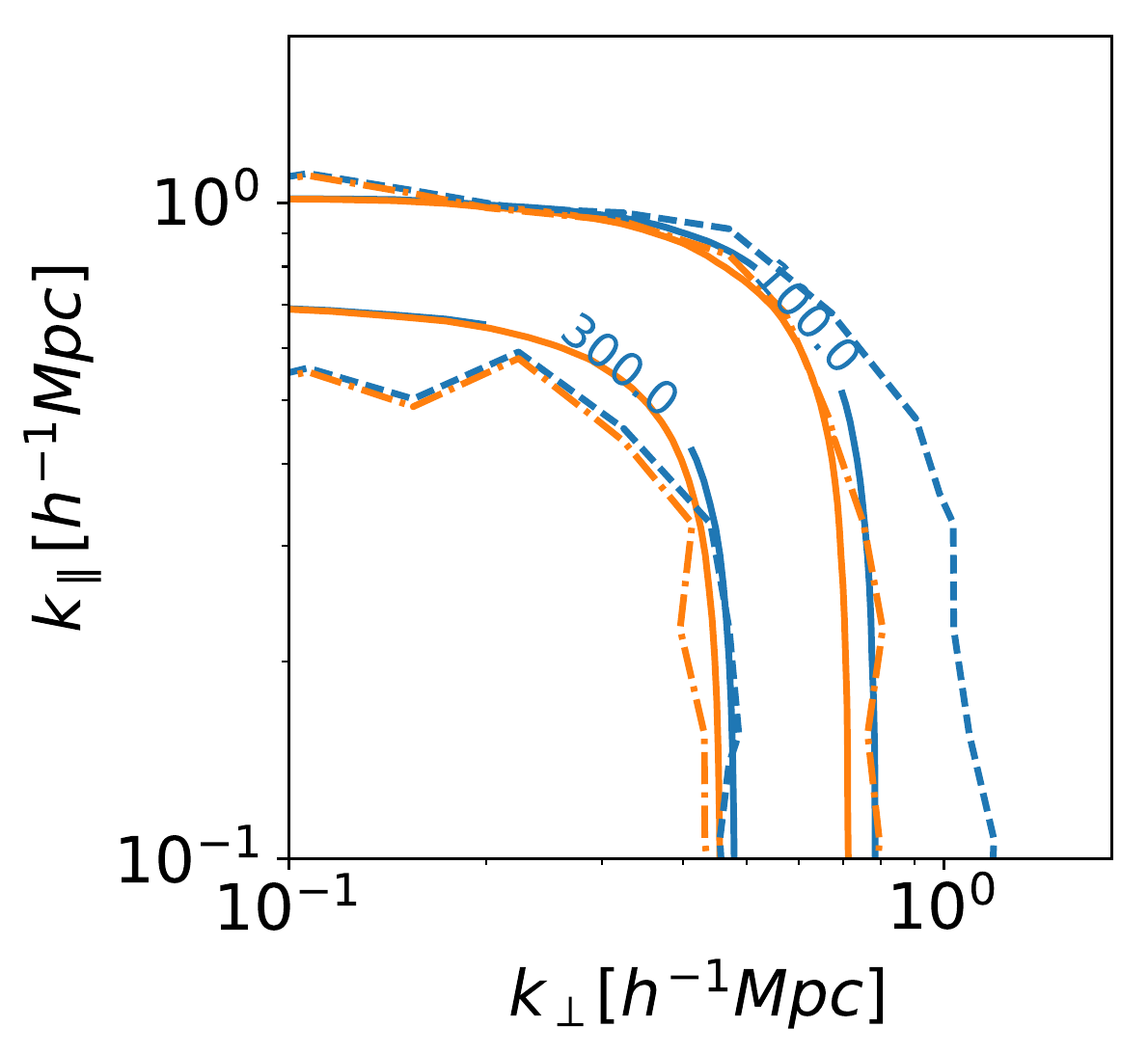}&
    \includegraphics[width=0.3\linewidth]{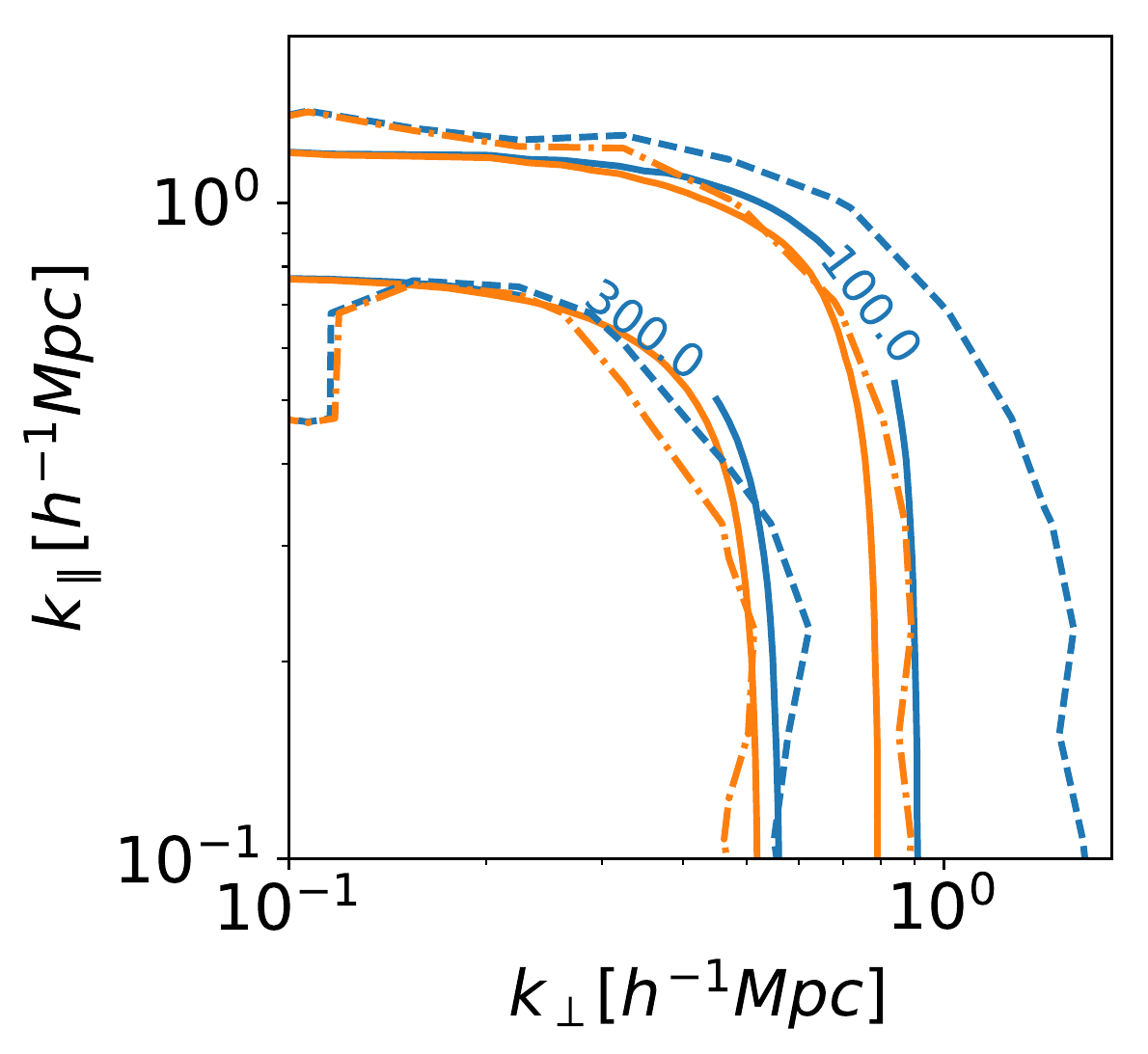}\\
  \end{tabular}
  \caption{
    Two dimensional power spectra of the \HI\ gas density, $P(k_{\perp},k_{\parallel})$ at z=3, 4, and 5 from left to right,  measured from the Osaka simulation. The abscissa and ordinate are the wave numbers perpendicular and parallel to the LoS.
The upper panels compare the power spectra in the real and redshift space, which are expected to be observed with different angular resolutions. 
Comparing spectra in real space with the one in redshift space, one can see that the spectra is elongated along the LoS due to the Kaiser effect, while on small scales the FoG effect is more prominent and the spectra are squashed.
Lower panels compare the simulated results with the theoretical predictions given by the TNS model with the best-fitting parameters given in Tables \ref{tab:best_params_bias} and \ref{tab:best_params_vsigma}. 
    \label{fig:Pkmu_mock}}
\end{figure*}

Figure~\ref{fig:Pkmu_mock} shows the 2D \HI\ power spectra $P(k_{\parallel},k_{\perp})$ in redshift space and real space. We see that the $P(k_{\parallel},k_{\perp})$ is elongated along the LoS direction.
While the elongation is more significant on larger scales due to the Kaiser effect, on smaller scales, the power spectrum shrinks in redshift space, which is caused by the non-linear velocities. The transition from elongation to squashing occurs at $k\simeq 1\,\hmpci$. Figure~\ref{fig:Pkmu_mock} also shows the spectra with an angular resolution of SKA-like observation in the interferometer and the single-dish observation modes. Since the single-dish observation has low angular resolution, the small-scale fluctuations in the transverse direction are considerably smoothed out. Conversely, the power spectrum for the interferometer map has a negligible effect of smoothing.

\subsection{RSD model, Kaiser Effect, and Fingers of God}
\label{ssec:model-RSD}
In this section, we describe the theoretical models for the anisotropic power spectrum of the \HI\ gas distribution in redshift space.
\HI\ gas traced by 21cm line is also affected by the peculiar motion of the gas clouds exactly in the same way as the galaxy doppler shift.
In the limit of linear theory for the velocity, Kaiser formula gives
\begin{equation}
	\label{eq:KaiserLin}
  P^{(s),{\rm Kaiser}}_{\HI}(k,\mu) = b^2_{\HI} (1+\beta \mu^2)^2 P_{\rm dm}^{\rm lin}(k),
\end{equation}
where $\mu$ is cosine of the angle between LoS and wavenumber vector $\bs{k}$, $b_{\HI}$ is the \HI\ bias, which we assume to be constant at first, 
and $\beta=f/b$ is the linear growth rate divided by the bias. Phenomenologically, the Fingers-of-God effect \rev{\RA{\citep{Jackson+1972}}} can be included as
\begin{equation}
	\label{eq:lin_fog}
    P^{(s)}_{\HI}(k,\mu)
    =
    {\rm DFoG}[k\mu f\sigma_v]\, P^{(s), {\rm Kaiser}}_{\HI}(k,\mu),
\end{equation}
where the prefactor DFoG represents the effect of damping given by either Gaussian or 
Lorentzian function in the literature \rev{\RA{\citep{Peacock+1994,Park+1994,Ballinger+1996,Magira+2000}}},
\begin{equation}
  {\rm DFoG}[x] = \left\{
    \begin{array}{ll}
      \exp(-x^2) & {\rm Gaussian}\\
      \displaystyle{\frac{1}{1+x^2}} & {\rm Lorentzian.} \\
    \end{array}
  \right.
\end{equation}
The velocity dispersion $\sigma_v$ can be given by either linear theory or just treated as free parameters later on.
If we focus on smaller scales, we may need to model the non-linearity of the density and
velocity fields, which can be given as
\begin{equation}
\label{eq:nl_kaiser}
	P^{(s),{\rm Kaiser}}_{\HI}(k,\mu)
    =
    b^2 \left(
    	P_{\delta\delta}(k) + 2\beta \mu^2 P_{\delta \theta}(k) + \beta^2 \mu^4 P_{\theta\theta}(k)
    \right),
\end{equation}where $\theta$ represents the divergence of velocity for \HI\ gas  and we assume no velocity bias \rev{\RA{\citep{Scoccimarro+2004}}}.
Here $P_{\delta\delta}=P_{\rm dm}$ is the dark matter power spectrum, $P_{\theta\theta}$ is velocity divergence power spectrum and 
$P_{\delta\theta}$ is the cross-power spectrum. Note that  Eq.~(\ref{eq:nl_kaiser}) is reduced 
to Eq.~(\ref{eq:KaiserLin}) in the limit of linear theory.
For further non-linear correction, we consider the additional terms in \cite{TNS2010}: 
\begin{equation}
\label{eq:nl_TNS}
  P^{(s)}_{\HI}(k,\mu)
  =
  {\rm DFoG}[k\mu f \sigma_v] 
  \left( P^{(s) {\rm Kaiser}}_{\HI}(k,\mu) 
  + b_{\HI}^3 A(k,\mu) 
  + b_{\HI}^4 B(k,\mu) \right), 
\end{equation}
where $A$ and $B$ terms are introduced so that the modulation of BAO
is readily accounted for at relatively large scales in quasi-
nonlinear regimes. In this paper, we compute $P^{(s),{\rm Kaiser}}_{\HI}$ using a publicly available code, \texttt{RegPT} \citep{Taruya+:2012} up to 2-loop order for $P_{\delta\delta}, P_{\delta\theta}$ and $P_{\theta\theta}$.  As an alternative model, we replace the linear power spectrum
in Eq.~(\ref{eq:KaiserLin}) with the full non-linear power spectrum $P^{\rm NL}(k)$, obtained by the fitting formula by \citet{Takahashi+:2012}.
Furthermore, we consider the angular resolution of the intensity mapping survey in the anisotropic power spectra. Since the density fluctuations on the direction perpendicular to the line-of-sight are observed convolved with the antenna beam function, the observed anisotropic power spectrum can be written as
\begin{equation}
  P^{(s)}_{\HI,{\rm obs}}(k,\mu)
  =
  W_{{\rm beam}}^2(k,\mu)P^{(s)}_{\HI}(k,\mu),
\end{equation}
where we assume that $W_{\rm beam}$ is Gaussian function
\begin{equation}
  W_{\rm beam}
  =  \exp\left(-\frac{k^2(1-\mu^2)\:\sigma^2_{{\rm smooth}}}{2}\right).
\end{equation}
$\sigma_{{\rm smooth}}$ is the comoving scale corresponding to the angular resolution at observed 21 cm redshift.

In order to compare the model with the simulation results, it is useful to expand the anisotropic power spectrum in a Legendre polynomial series,
\begin{equation}
  P^{(s)}_l(k)
  =
  \frac{2l+1}{2}
  \int_{-1}^1 \! {\rm d}\mu\, P^{(s)}(k,\mu) {\mathcal L}_l(\mu),
\end{equation}
where $l=0, 2$ or $4$, and ${\mathcal L}_l$ is the $l$-th Legendre polynomial.

\begin{figure*}
	\centering
    \begin{tabular}{cc}
        \includegraphics[width=0.45\linewidth]{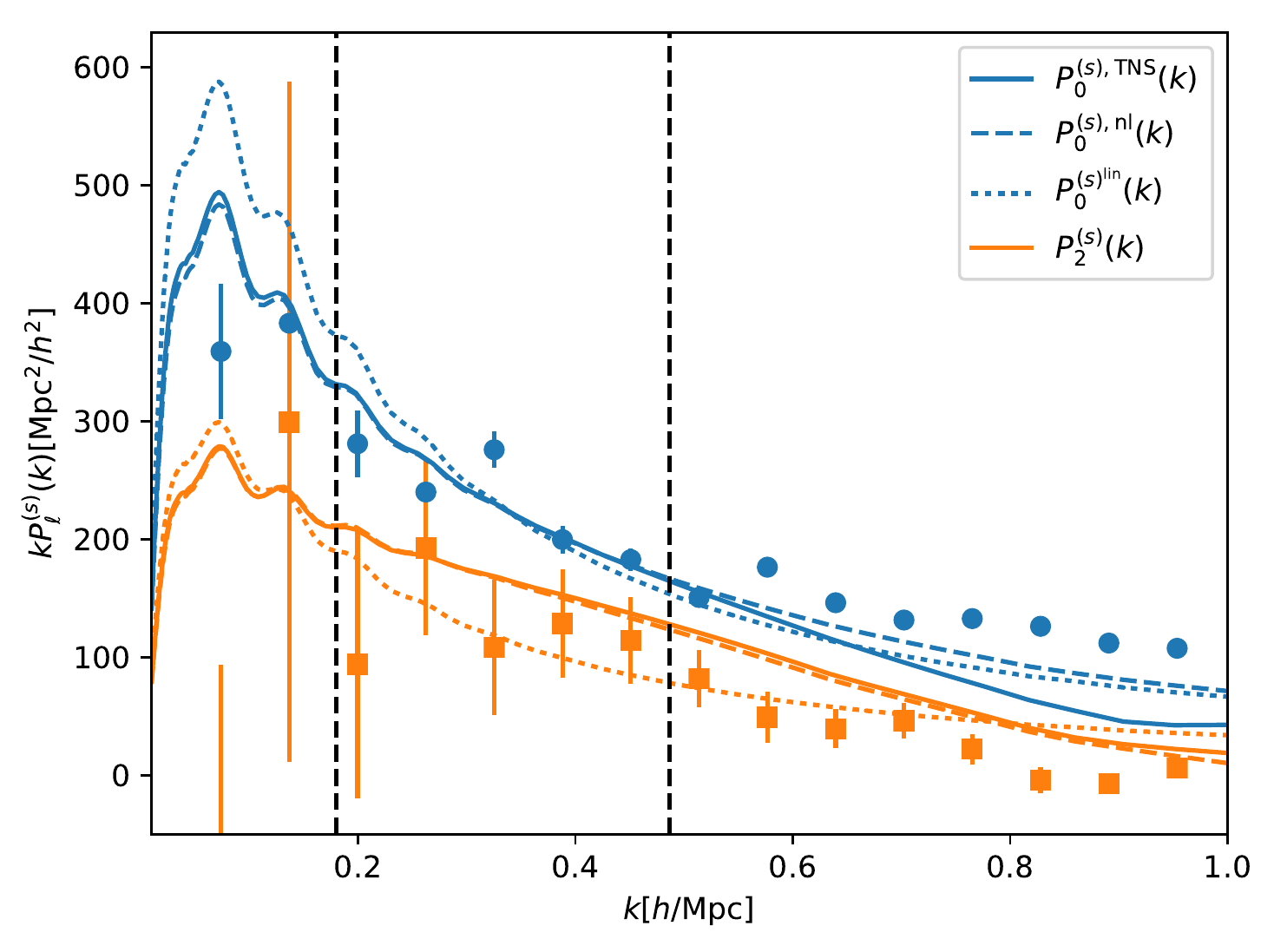}&
 		\includegraphics[width=0.45\linewidth]{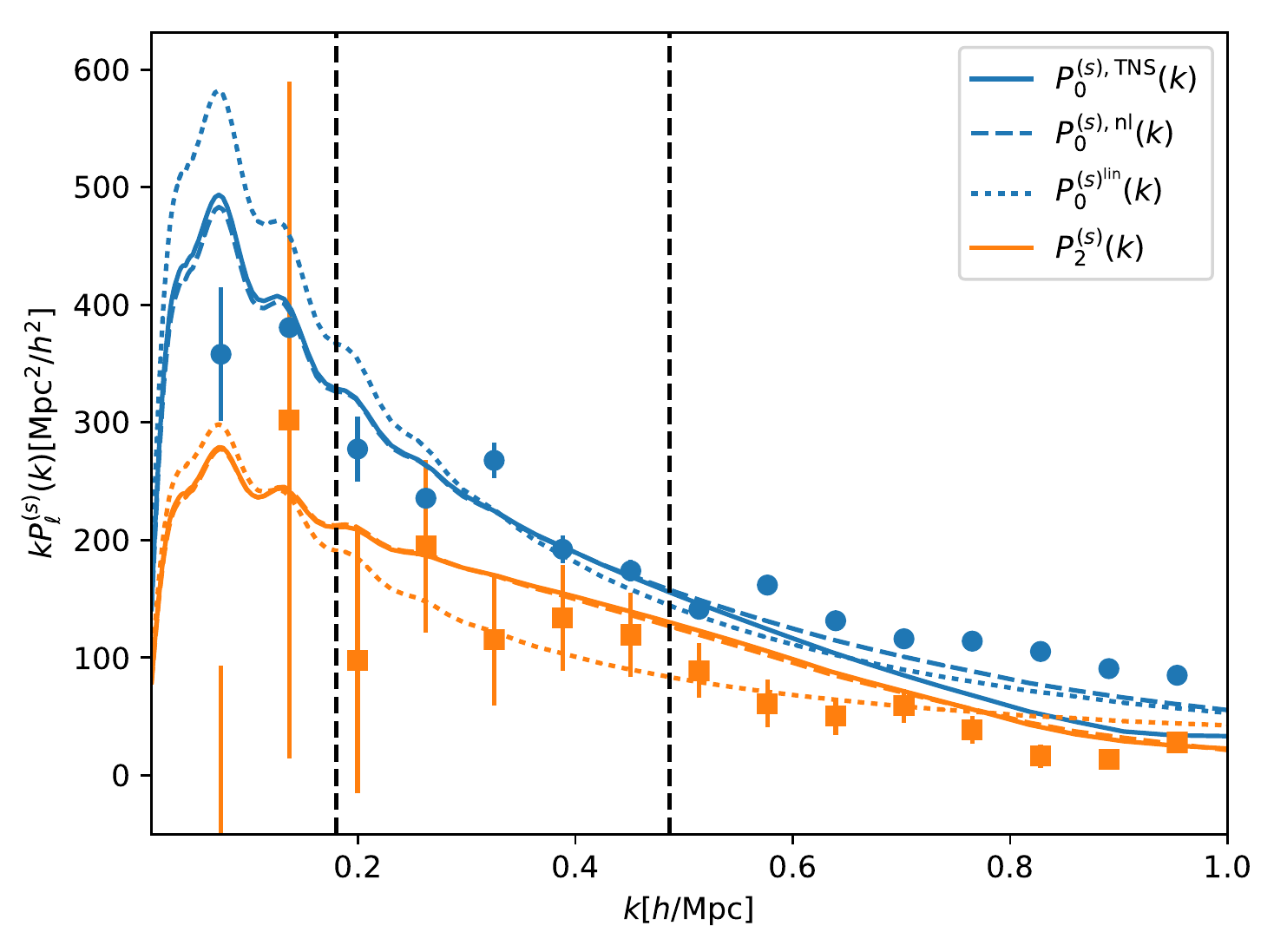}
    \end{tabular}
  	\caption{
   {\it Left panel}: Legendre expansion of anisotropic power spectra.  
   The blue upper set of lines and symbols are monopole, and orange lower set of lines are quadrupole. Symbols are measured from the Osaka simulation, and the curves are the best-fitting models for $0.18<k<k_{\rm max}$, using the TNS (solid), non-linear empirical (dashed) and linear theory (dotted), respectively. The vertical dashed line shows the scale of 0.18 and $k_{\rm max}$.  {\it Right panel}: Same as the left panel, but with the angular resolution assuming the angular resolution of the interferometer mode. \rev{{Error bars are the standard deviation of $P^{(s)}_\ell(k)$ in each bin divided by the number of modes, $\sigma/\sqrt{N_k}$.}}
    \label{fig:Pell}}
\end{figure*}

For the simulation data, we calculate the multipole \HI\ power spectra $P^{(s)}_l(k)$ as,
\begin{equation}
	\label{eq:def_pkmu}
	\hat{P}_{l}^{(s)}(k_i) = \frac{2l+1}{2} \sum_{j, k_j \in k_i}^{N_k} \Re{[\delta_{\HI}(k_j) \delta_{\HI}^*(k_j)]}\: {\mathcal L}_l(\mu_j)\: \Delta \mu_k, 
\end{equation}
where $\mu=k_{\parallel}/k, \Delta \mu_k=2/N_k$ and $N_k$  is the number of modes within the $j$-th wavenumber bin. 
We then find the best-fitting parameters with the following likelihood: 
\begin{equation}
\label{eq:Likeli_pleg}
   {\mathcal L} \propto \exp\left( - \frac{\chi_{l=0}^2+\chi_{l=2}^2}{2}\right),
\end{equation}
where $\chi_l$ is the chi-square for the $l$-th moment,
\begin{equation}
\label{eq:chisq_pleg}
   \chi_l^2=\sum_i^{k_i<k_{\rm max}} 
   \frac{\left[ \hat{P}_{l,i}^{(s)}-P_l^{(s)}(k_i) \right]^2}{\sigma_i^2}.
\end{equation}
We fit the results with three models for $0.18<k<k_{\rm max}$. 
We set the linear {\HI} bias $b_0$ and velocity dispersion $\sigma_v$ (see Eq.\,(\ref{eq:lin_fog})) as free parameters and fit within the range of $0.1\leq b_0 \leq 5$ and $0\leq \sigma_v \leq 5\,\sigma_{\rm lin}$, where $\sigma_{\rm lin}$ is the velocity dispersion predicted from the linear perturbation theory.
As we only use the single realisation of simulation, the large-scale fluctuation is highly affected by the cosmic variance. We find that the dark matter power spectra at $k<0.18\,\hmpci$ of our simulation has significantly smaller amplitude compared to the theoretical prediction of the matter power spectrum and thus a lower \HI\ bias is favoured. 
(Note that the wave number corresponding to the box-size is $k_{\rm min}=2\pi/L_{\rm box}=0.074 h^{-1}$Mpc.)
Therefore we decide to remove the two largest modes from the fitting. Note that the bias measurement in Eq.\,(\ref{eq:bias_cross}) does not suffer from this effect because the amplitude suppression appear both in dark matter and \HI\ and they are cancelled out. The maximum wavenumber $k_{\rm max}$ is again given by Eq.\,(\ref{eq:k_max}).

Figure \ref{fig:Pell} compares the measured power spectra and the following models with best-fitting parameters: 
linear model (Eqs.\,(\ref{eq:KaiserLin}) \& (\ref{eq:lin_fog})), alternative model 
(i.e. the linear power spectrum $P_{\rm dm}^{\rm lin}(k)$ in Eq.\,(\ref{eq:KaiserLin}) is replaced with the non-linear power spectrum), and TNS model (Eq.~(\ref{eq:nl_TNS})).
We find that the alternative model and the TNS model agree well with the measured {\HI} power spectra.
The chi-square values $(\chi_0^2+\chi_2^2)$ for the linear model, alternative model, and TNS model are 28, 17, and 17,  respectively; that is, the fitting by the alternative model and TNS model are improved by $\Delta\chi^2=12$, suggesting that these two models are better than the linear model. 
Since $P_{\rm dm}^{\rm lin}(k)$ in Eq.\,(\ref{eq:KaiserLin}) is simply derived from the linear theory, 
the linear model does not fully consider the non-linearity. 
Therefore the best-fitting value of velocity dispersion $\sigma_v$ for the linear model is smaller than others. 
The best-fitting parameters for each model are shown in Tables~\ref{tab:best_params_bias} and \ref{tab:best_params_vsigma}.

\section{Implications}
\label{sec:results}
In this section, we compare the \HI\ biases measured from the real space power spectra between Illustris and Osaka simulations which are based on different prescriptions of astrophysical effects.
We also discuss the bias and velocity dispersion parameters derived from the redshift space observation to understand the \HI\ clustering properties.

\subsection{Comparison between Illustris and Osaka simulations}
\label{ssec:astro}
Here we compare the two different simulations in terms of \HI\ bias behaviour. On large scales, the overall behaviour of the two simulations is similar to each other. 
Although the two simulations have different astrophysical effects as described in Section~\ref{sec:simulation}, those effects are only seen on small scales and on larger scales of $k<k_{\rm max}$ defined in Eq.~(\ref{eq:k_max}), the effects do not significantly affect the values of \HI\ bias both for constant and scale dependent terms. This suggests that details of the astrophysical effects in the simulation are not important for measuring the scale dependent bias and that the future analysis of measuring the BAO scales are robust against the baryonic effects.

\subsection{Bias and velocity dispersion}
\label{ssec:bias-velocity}

Here we explore the linear bias and velocity dispersion parameters (see Eq.~\ref{eq:lin_fog}) simultaneously fitted to the redshift space power spectra, keeping the cosmological parameters fixed to the values used in the simulations.
The detailed procedure of estimating the best-fitting parameters are described in Section~\ref{ssec:model-RSD}.

\begin{figure}
\centering
\begin{tabular}{c}
  \includegraphics[width=\linewidth]{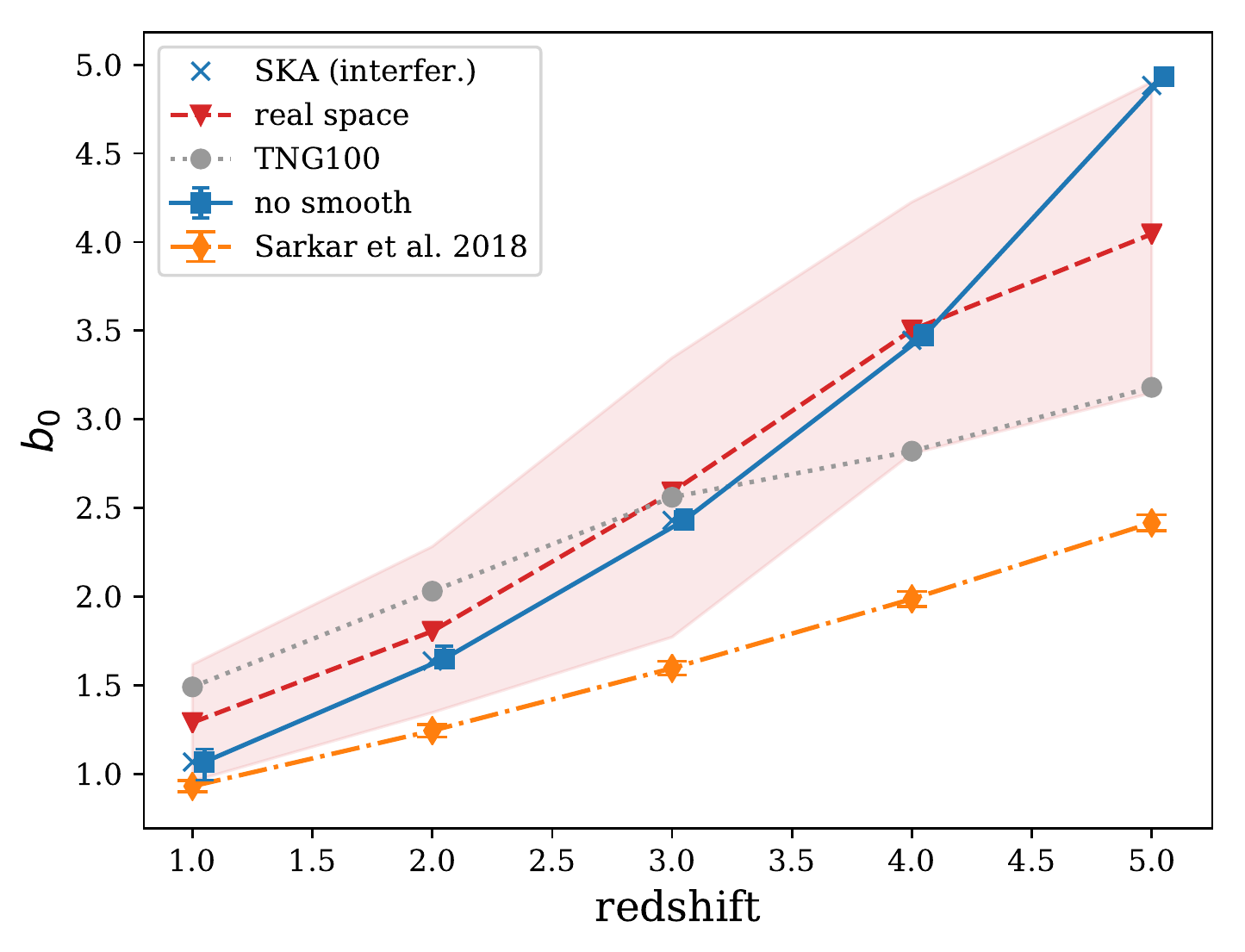}\\
  \includegraphics[width=\linewidth]{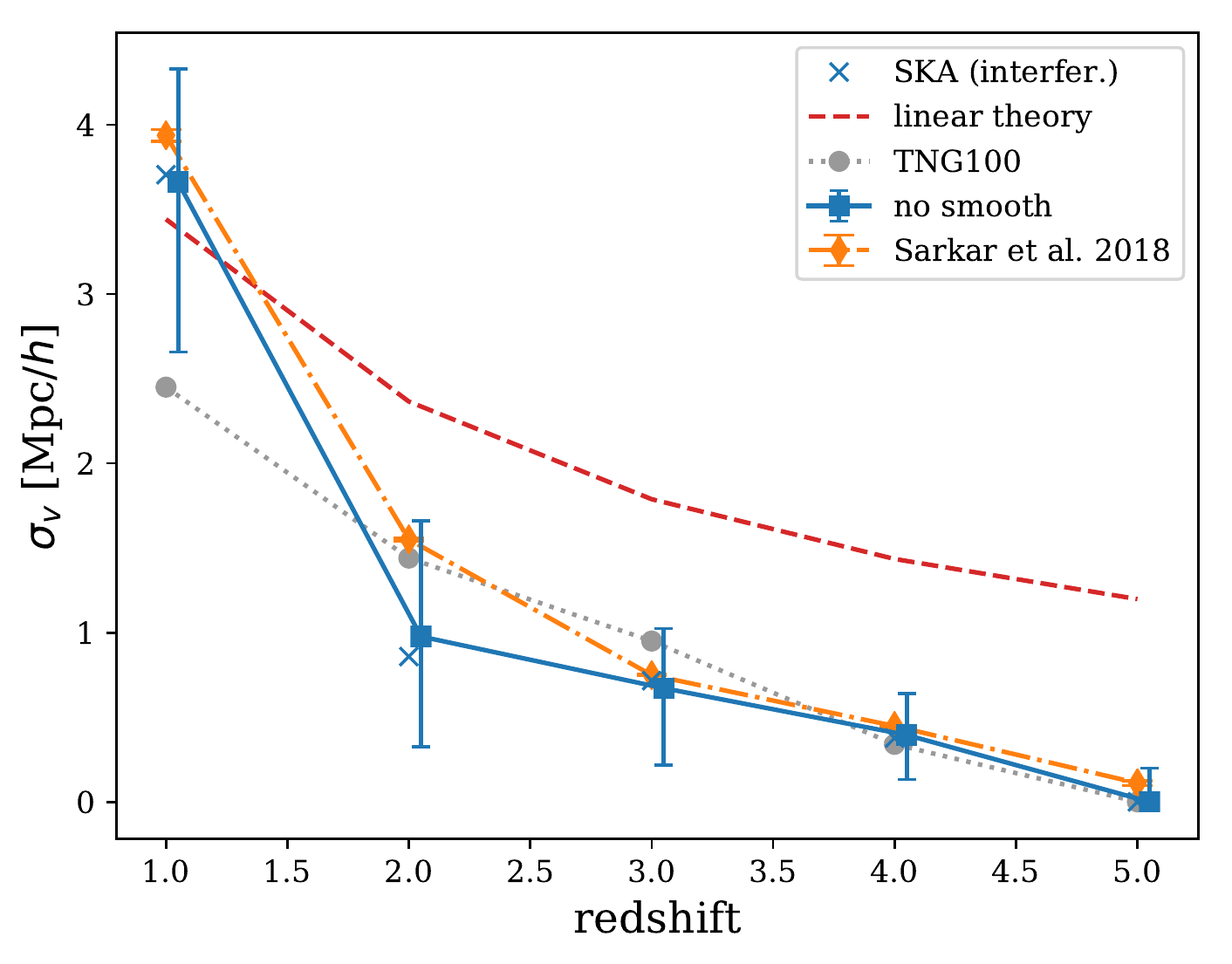}\\
  \end{tabular}
  \caption{
	(\textit{Top}:) Best-fitting values of $b_{\HI}$ and those obtained for the real space power spectrum $b_0$ defined in Eq.~(\ref{eq:bias_polynomial}) (dashed-line with triangle). The shaded region is a 1-sigma uncertainty on $b_0$ measurement.
   (\textit{Bottom}:) Best-fitting values of $\sigma_v$ compared with those predicted from linear perturbation theory (dashed-line). 
   For both panels, the square symbols with error bars and the blue crosses correspond to the fitting results with the TNS model to the non-smoothed and high-angular resolution data, respectively. For both panels, also shown with orange dashed-dotted-line and the diamond symbols is the best-fitting model of B1 (HC)  of \citet{Sarkar+2018a}. 
   \rev{{The results from TNG simulation \citep{TNG+2018} are shown with grey dotted lines with circles in both panels.}}
  \label{fig:b0_sigv}}
\end{figure}

Figure~\ref{fig:b0_sigv} shows the redshift dependence of bias and velocity dispersion of \HI\ gas obtained by fitting the Legendre polynomial expanded data with the TNS model. 
The best-fitting models are compared with the real space measurement of the bias or the prediction from linear perturbation theory for the velocity dispersion. 
In the top panel, we find that the constant bias is systematically smaller than that obtained by the direct measurement using Eq.~(\ref{eq:bias_polynomial}).  With a low angular resolution, the bias is significantly underestimated though it is still consistent with the direct measurement within 1-$\sigma$ error. 
In the bottom panel of Fig.~\ref{fig:b0_sigv}, we find that the velocity dispersion of \HI\ gas is marginally consistent with the linear theory prediction at $z\sim 1$, but it is systematically smaller than the linear theory at $z>2$, although the uncertainties are still large. 

It is not feasible to estimate the two dimensional anisotropic power spectrum, because the angular resolution of single-dish mode is comparable or worse than the box-size of our simulation, and the transverse fluctuation is fully smoothed out.  Thus we cannot measure the bias parameters for single-dish mode and need simulations with larger box-size. 
Both bias and velocity dispersion are underestimated even for the no-smooth case, and we need to develop in the future a more sophisticated model for the non-linear {\HI} power spectra which takes the coarse angular resolution into account.


\newlength{\tabh}
\setlength{\tabh}{0.5cm}
\begin{table*}
  \begin{tabular*}{\linewidth}{@{\extracolsep{\fill}}c|cc|cc|cc}\hline \hline
  redshift&TNS&TNS&non-linear&non-linear&linear&linear\\ 
  ~ & no smooth & high res. & no smooth & high res. & no smooth & high res. \\ \hline
\parbox[c][\tabh][c]{0cm}{}1&\val{1.07}{-0.10}{+0.07}&\val{1.07}{-0.10}{+0.07}
			&\val{1.03}{-0.10}{+0.07}&\val{1.03}{-0.10}{+0.07}
			&\val{1.12}{-0.07}{+0.07}&\val{1.12}{-0.07}{+0.07}\\
\parbox[c][\tabh][c]{0cm}{}2&\val{1.65}{-0.05}{+0.07}&\val{1.64}{-0.04}{+0.08}
			&\val{1.61}{-0.07}{+0.06}&\val{1.60}{-0.06}{+0.07}
			&\val{1.85}{-0.03}{+0.06}&\val{1.85}{-0.04}{+0.06}\\
\parbox[c][\tabh][c]{0cm}{}3&\val{2.43}{-0.05}{+0.06}&\val{2.43}{-0.06}{+0.05}
			&\val{2.39}{-0.06}{+0.06}&\val{2.39}{-0.07}{+0.05}
			&\val{2.69}{-0.04}{+0.05}&\val{2.68}{-0.04}{+0.05}\\
\parbox[c][\tabh][c]{0cm}{}4&\val{3.47}{-0.04}{+0.06}&\val{3.45}{-0.04}{+0.06}
			&\val{3.45}{-0.06}{+0.05}&\val{3.42}{-0.06}{+0.04}
			&\val{3.82}{-0.04}{+0.04}&\val{3.79}{-0.03}{+0.05}\\
\parbox[c][\tabh][c]{0cm}{}5&\val{4.93}{-0.03}{+0.05}&\val{4.88}{-0.04}{+0.04}
			&\val{4.90}{-0.04}{+0.04}&\val{4.83}{-0.03}{+0.05}
			&\val{5.35}{-0.04}{+0.05}&\val{5.32}{-0.05}{+0.04}\\
\hline\hline
  \end{tabular*}
  \caption{
  Best-fitting parameters for the constant bias $b_{\HI}$ in the non-smoothed case and high-angular resolution case. Superscript and subscript are upper and lower 68 percentiles.
  \label{tab:best_params_bias}}
\end{table*}
\setlength{\tabh}{0.5cm}
\begin{table*}
\centering
  \begin{tabular*}{\linewidth}{@{\extracolsep{\fill}}c|cc|cc|cc}\hline \hline
  redshift&TNS&TNS&non-linear&non-linear&linear&linear\\ 
  ~ & no smooth & high res. & no smooth & high res. & no smooth & high res. \\ \hline
\parbox[c][\tabh][c]{0cm}{}1&\val{3.66}{-1.01}{+0.67}&\val{3.71}{-1.05}{+0.62}
			&\val{3.96}{-0.88}{+0.61}&\val{3.96}{-0.89}{+0.60}
			&\val{2.02}{-1.38}{+0.89}&\val{2.02}{-1.38}{+0.89}\\
\parbox[c][\tabh][c]{0cm}{}2&\val{0.98}{-0.65}{+0.68}&\val{0.86}{-0.54}{+0.79}
			&\val{1.30}{-0.89}{+0.52}&\val{1.24}{-0.84}{+0.56}
			&\val{0.00}{}{+0.97}&\val{0.00}{}{+0.97}\\
\parbox[c][\tabh][c]{0cm}{}3&\val{0.67}{-0.45}{+0.35}&\val{0.72}{-0.50}{+0.30}
			&\val{0.87}{-0.57}{+0.27}&\val{0.90}{-0.60}{+0.24}
			&\val{0.00}{}{+0.49}&\val{0.00}{}{+0.49}\\
\parbox[c][\tabh][c]{0cm}{}4&\val{0.40}{-0.26}{+0.25}&\val{0.38}{-0.26}{+0.24}
			&\val{0.56}{-0.37}{+0.17}&\val{0.54}{-0.37}{+0.16}
			&\val{0.00}{}{+0.26}&\val{0.00}{}{+0.26}\\
\parbox[c][\tabh][c]{0cm}{}5&\val{0.00}{}{+0.20}&\val{0.00}{}{+0.19}
			&\val{0.00}{}{+0.21}&\val{0.00}{}{+0.20}
			&\val{0.00}{}{+0.13}&\val{0.00}{}{+0.13}\\
\hline\hline
\end{tabular*}
  \caption{
  Same as Table \ref{tab:best_params_bias} but for $\sigma_v$ parameter.
  \label{tab:best_params_vsigma}}
\end{table*}

\rev{{In Section\,\ref{sec:introduction}, we explained the reason why we primarily compare the Osaka simulation with Illustris-3 simulation. Here we briefly discuss the comparison with the IllustrisTNG simulation \citep{TNG+2018} which is a magneto-hydrodynamic simulation  with box size 75\,{$\himpc$} and mean baryonic particle mass being $1.4\times 10^6\,\Msun$ for TNG100-1. 
}
\RA{The box-size and the particle resolution of TNG100-1 are same as that of Illustris-1.} 
{
In TNG100-1, we find that the {\HI} bias measured in real space is larger at low redshifts and smaller at high redshifts than in Osaka simulation.  
In other words, the redshift evolution of bias in Osaka \& Illustris-3 simulation is stronger than in TNG100-1.
We also compare the velocity dispersion parameter measured in redshift space with TNG100-1, and find that the results are consistent among the simulations within 1-$\sigma$ statistical error at $z>1$.
}}

\section{Summary and Conclusions}
\label{sec:summary}
Future 21 cm surveys will reveal the three dimensional distribution of neutral hydrogen gas over cosmological scales, which will potentially be a new probe of the large-scale structure of the Universe.
In this paper, we explore the properties of \HI\ clustering using two different set of cosmological hydrodynamic simulations, the Illustris and the Osaka simulations that include nonlinear baryonic effects of star formation and feedback. 

We first measure the scale and redshift dependences of \HI\ bias in real space by taking the ratio of power spectra of {\HI}--dark matter cross correlation and dark matter auto correlation. Fitting with the constant plus linearly-scaled bias with $k$,
\rev{{we find that the \HI\ bias monotonically increases with redshift for both simulations. This result is consistent with the Illustris-TNG simulation \citep{TNG+2018}, but the redshift evolution is stronger in the  Illustris-1\,\&\,3 than in TNG100-1.}}  We also find that the \HI\ bias shows a significant scale dependence at $z>4$ up to the scales where the perturbation theory holds, but it is consistent with being constant at $z\leq 3$.  If we limit our analysis to the large scales of $k<0.25\,\hmpci$, we find no evidence of scale dependence at $1<z<5$.  In both cases, the best-fitting bias parameters are fairly consistent between Illustris and Osaka simulations, which implies that the scale dependence of \HI\ bias on large scales is not sensitive to the details of the small-scale astrophysics. This means that, as far as we use the large scale modes, the cosmological analysis such as the determination of BAO scale is unlikely to be affected by the astrophysical uncertainties of feedback on small scales. 
However, at the same time, if one use the data more aggressively up to higher $k$, we certainly need accurate  knowledge on the astrophysical effects such as supernova or AGN feedback. We leave more detailed and thorough investigation of the astrophysical impact of feedback on the \HI\ power spectrum as a future work. \rev{\RA{Further discussion on the evolution of $\Omega_{\rm HI}$ is given in the appendix.}}

We then measure the redshift space distortion using the anisotropic two dimensional power spectrum. We jointly fit the monopole and quadrupole of the Legendre expanded power spectra including the peculiar velocity effect to the models widely applied for galaxy redshift surveys, with the free parameters of bias $b_{\HI}$ and velocity dispersion $\sigma_v$. We note that, since we only have two simulations, the cosmic variance largely affects the amplitude of large scale fluctuations. Therefore, we fit the data only in the range of $0.18<k<k_{\rm max}$ where $k_{\rm max}$ is given by Eq.~(\ref{eq:k_max}). We find that the measured bias parameter in redshift space is consistent with the one directly measured in real space from the ratio of power spectra. We also find that the velocity dispersion of \HI\ gas is systematically below the prediction from linear perturbation theory but marginally consistent with the prediction.

Compared with the previous work by \citet{Sarkar+2018a}, we find significant disagreement on the values of \HI\ bias for the entire redshift range, which may mainly arise from the prescription of the \HI\ gas assignment to the dark matter halos in an $N$-body simulation by \citet{Sarkar+2018a}. On the other hand, the best-fitting values of our velocity dispersion are fairly consistent with the previous work within the statistical error of the single box simulation. Although the simulations used in this paper solve baryonic distribution and hydrogen ionization process in a more realistic manner, more detailed analysis will be required to fully understand the discrepancy. 


In this paper, we also introduced a new empirical model for RSD. The model is a simple replacement of linear power spectrum $P^{\rm lin}(k) \rightarrow P^{\rm NL}(k)$ in the Kaiser formula with the FoG prefactor. This model is consistent with the full TNS model on scales $k<k_{\rm max}$, and it gives a better fit of monopole at $k>k_{\rm max}$ but slightly off from the data for quadrupole on those scales.

We construct a mock simulated data assuming that the 21 cm line is observed by future SKA-like survey for both interferometer and single-dish modes. We find that, for single-dish observation (i.e. low angular resolution observation), the models systematically underestimate the bias parameter and velocity dispersion. It will require a model in which the coarse angular resolution has been taken into account.

In this paper we have limited ourselves to the discussion of  the clustering properties of \HI\ gas, however it is straightforward to extend our analysis to the cosmological parameter recovery, such as the growth rate $f$ or dark energy parameters $w$ or $\Omega_{\rm DE}$. We leave these analysis to our future work, in which we will employ a larger number of hydrodynamic simulations with larger box-sizes. 

\appendix
\section{Neutral hydrogen abundance}
\label{sec:Omega}

\begin{figure}
\centering
  \includegraphics[width=\linewidth]{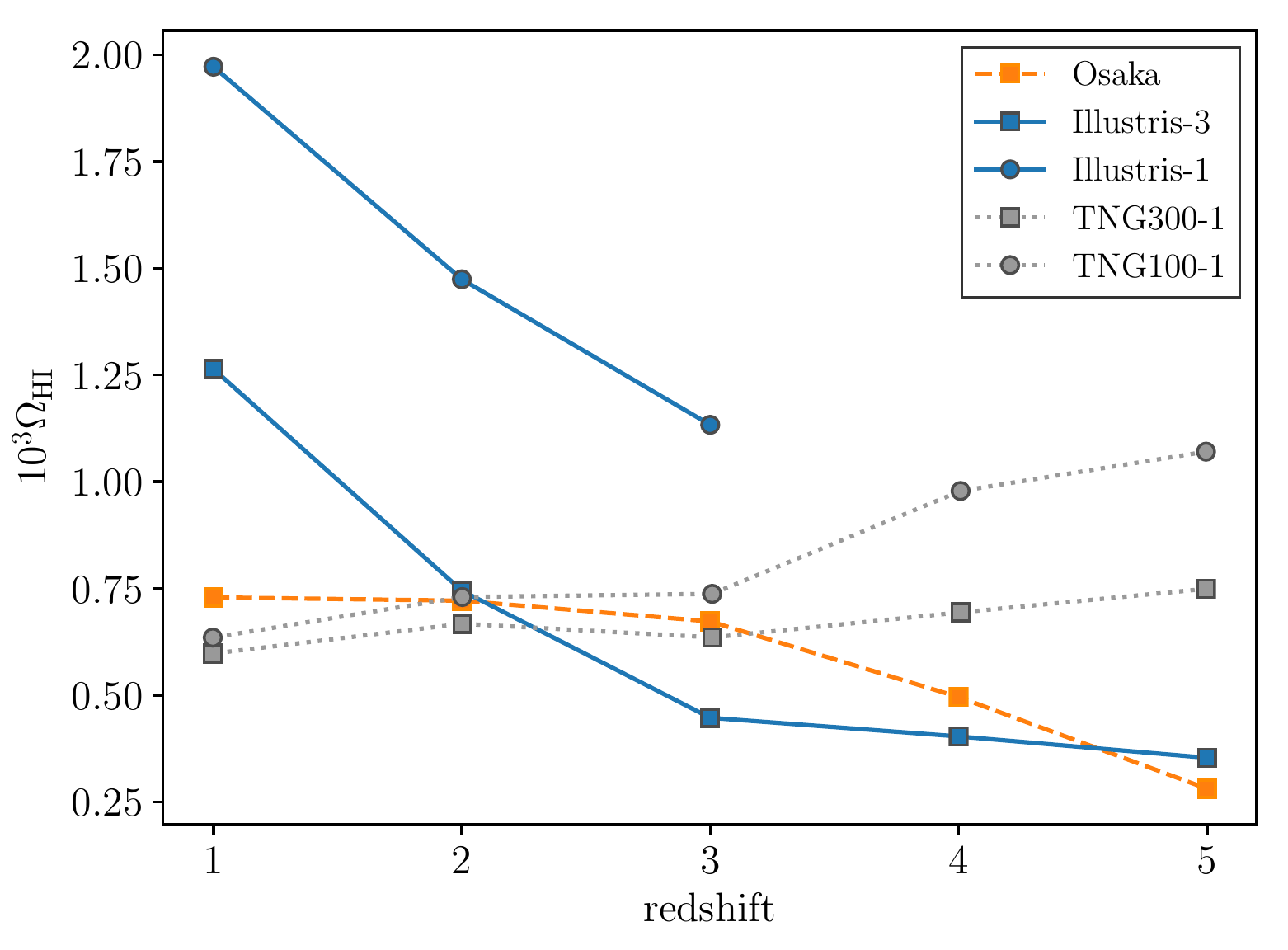}
  \caption{
  \rev{\RA{{\HI} density parameter $\Omega_{\HI}\equiv \rho_{\HI}(z)/\rho_c(z=0)$ multiplied by $10^3$ as a function of redshift.  The dashed orange line with square symbols is measured from the Osaka simulation. The solid blue line with square and circle symbols correspond to the Illustris-3 and Illustris-1, respectively. The dotted line with square and circle symbols are measured by TNG \citep{TNG+2018}, and their box sizes are  $300\,\himpc$ and $100\,\himpc$ on a side, respectively.}} 
  \label{fig:omega_HI}}
\end{figure}

\rev{{It would be useful to check the global evolution of {\HI} density over cosmic time in order to understand the subtle discrepancies of {\HI} bias in different simulations.
In Fig.\,\ref{fig:omega_HI}, we compare $\Omega_{\HI}(z)\equiv \rho_{\HI}(z)/\rho_c(z$=$0)$ from different simulations, and find that the Illustris-1 has about twice higher $\Omega_{\HI}$ than Illustris-3 due to its higher resolution. 
This discrepancy can be fully explained by the minimum dark matter halo mass resolved in each simulation. The higher resolution simulation, Illustris-1 can resolve the dark matter halos down to $\sim 10^8\,\himsun$, while the Illustris-3 cannot resolve the halos with $\lesssim 10^{10}\,\himsun$. We have checked this by plotting the halo mass functions from both simulations using the publicly available data. Since most of the {\HI} resides in dark matter halos, the minimum dark matter halo mass that can be resolved in each simulation is directly reflected in the total amount of {\HI} gas.
On the other hand, the results from two Illustris-TNG simulations with different resolution (TNG100-1 \& TNG300-1) seem to converge well as shown in Fig.\,\ref{fig:omega_HI}.
The results of Illustris-1, 3 and TNG simulations are different at most by a factor 3, which can be ascribed to different efficiencies of SN \& AGN feedback models \citep{Pillepich2018, Weinberger+2018}. 
The feedback models in the TNG simulations were tuned to make them more effective than the original Illustris simulations, thereby suppressing the overabundant galaxies at both high- and low-mass end of the galaxy stellar mass function. 
As shown in Fig.\,4 of \citep{Pillepich2018}, the gas fraction in low-mass halos are significantly lower in the TNG simulation than in the original Illustris simulation at $z=0$, which is also reflected in the lower $\Omega_{\rm HI}$ in the TNG at $z<2$.}}

\section*{Acknowledgments}
We would like to thank Naoshi Sugiyama, Hiroyuki Tashiro, Atsushi Taruya, Fabian Schmidt, and Dominik Schwarz for useful discussions and anonymous referee to improve our manuscript.
AN is in part supported by JSPS KAKENHI Grant Number JP16H01096.
KH is in part supported by JSPS KAKENHI Grant Numbers JP17H01110 and JP18K03699.
IS and KN acknowledge the support by the JSPS KAKENHI Grant Number JP17H01111. 
KN acknowledges the travel support from the Kavli IPMU, World Premier Research Center Initiative (WPI), where part of this work was conducted. 
The Osaka simulations were performed on Cray XC30 at CfCA, National Astronomical Observatory of Japan, 
We also utilised the {\small OCTOPUS} at the Cybermedia Centre, Osaka University, as part of the HPCI system Research Project (hp180063).

\bibliographystyle{mn2e}
\bibliography{bibdata}

\end{document}